\documentclass[superscriptaddress,noshowpacs,showkeys,twocolumn,floatfix,aps,prb,reprint,footinbib,citeautoscript]{revtex4-1}
\usepackage[english]{babel}
\usepackage{fontenc}
\usepackage{graphicx}
\usepackage{amsmath}
\usepackage{epstopdf}
\usepackage{amssymb}
\usepackage{amsbsy}
\usepackage{amscd}
\usepackage{color}
\usepackage{bbm}
\usepackage{braket}
\usepackage{bm}
\usepackage{siunitx}
\usepackage[normalem]{ulem} 
\usepackage{verbatim}
\usepackage{url}
\usepackage[verbose,hypertexnames=false,bookmarksopenlevel=1,filecolor=blue,linkcolor=blue,citecolor=blue,urlcolor=blue,pdfstartview=FitH,bookmarksopen,bookmarksnumbered,colorlinks,plainpages=false,linktocpage]{hyperref}
\DeclareSymbolFont{mathbold}{OML}{cmm}{b}{it}
\DeclareMathSymbol{\bsigma}{\mathord}{mathbold}{27}
\graphicspath{{./figs/}}
\begin{document}
\title{In-plane magnetoelectric response in bilayer graphene}
\author{Michael Kammermeier}
\email{michael.kammermeier@vuw.ac.nz}
\affiliation{School of Chemical and Physical Sciences and MacDiarmid Institute
for Advanced Materials and Nanotechnology, Victoria University of Wellington,
P.O. Box 600, Wellington 6140, New Zealand}
\author{Paul Wenk}
\affiliation{Institute for Theoretical Physics, University of Regensburg, 93040
Regensburg, Germany}
\author{Ulrich Z\"ulicke}
\affiliation{School of Chemical and Physical Sciences and MacDiarmid Institute
for Advanced Materials and Nanotechnology, Victoria University of Wellington,
P.O. Box 600, Wellington 6140, New Zealand}
%%----------------------------------------------------------------------
\date{\today }
%----------------------------------------------------------------------
\begin{abstract}
A graphene bilayer shows an unusual magnetoelectric  response whose
magnitude is controlled by the valley-isospin density, making it possible to link
magnetoelectric behavior to valleytronics. Complementary to previous studies,
we consider the effect of static homogeneous electric and magnetic fields that
are oriented parallel to the bilayer's plane. Starting from a tight-binding
description and using quasi-degenerate perturbation theory, the low-energy
Hamiltonian is derived including all relevant magnetoelectric terms whose
prefactors are expressed in terms of tight-binding parameters. We confirm the
existence of an expected axion-type pseudoscalar term, which turns out to
have the same sign and about twice the magnitude of the previously obtained
out-of-plane counterpart. Additionally, small anisotropic corrections to the
magnetoelectric tensor are found that are fundamentally related to the skew
interlayer hopping parameter $\gamma_4$. We discuss possible ways to
identify magnetoelectric effects by distinctive features in the optical
conductivity.
\end{abstract}

\maketitle

%---------------------------------------------------------------------
\allowdisplaybreaks

%---------------------------------------------------------------------
\section{Introduction}

Although having been extensively studied for more than half a
century~\cite{ode70}, the magnetoelectric (ME) effect has recently returned
to the spotlight. The renewed attention is motivated by the discovery of new
material systems. The traditionally predominant focus on band insulators with
intrinsically broken space-inversion and time-reversal symmetries, known as
multiferroics~\cite{Fiebig2005,Ramesh2007,Spaldin2017,Fiebig2016}, has
expanded to topological insulators~\cite{Qi2008,Essin2009,Rosenberg2010,
Qi2011}, Weyl semimetals~\cite{Grushin2012,Zyuzin2012a,Zyuzin2012b,
Vazifeh2013}, and very recently to metals~\cite{Ma2015,Varjas2016,
Zhong2016,Thoele2018}. The latest finding of bilayer graphene being a ME
medium further extends the list by adding an example from the rapidly
expanding class of two-dimensional materials that have their own unique ME
properties~\cite{Zhang2011,Zuelicke2014,Winkler2015,Zuelicke2017}.

A particularly appealing feature of ME media is the possibility to manipulate
magnetic properties in a solid by electric fields and \textit{vice versa\/}. This
enables us to engineer device architectures with novel functionalities not
achievable with other materials~\cite{Ramesh2007,Fiebig2016,Spaldin2017}.
It also establishes an inspiring connection between ideas and methods from
condensed-matter physics, high-energy physics, and even
cosmology~\cite{Franz2008,Spaldin2017}. In a nutshell, the ME effect
becomes manifest in a mixing between electric and magnetic fields,
$\boldsymbol{\mathcal{E}}$ and $\mathbf{B}$, in the expansion of the free
energy $\mathcal{F}$. In leading-order, this coupling is linear in both fields and
described by the ME tensor $\alpha$, whose components are defined
as~\cite{Fiebig2005,Spaldin2008}
\begin{align}
\alpha_{ij}={}&-\frac{\partial\mathcal{F}}{\partial\mathcal{E}_i\partial B_j}
\: .
\end{align}
It is common practice to decompose the ME tensor as~\cite{Essin2010,
Spaldin2008}
\begin{align}
\alpha_{ij}={}&\alpha_\theta\;\delta_{ij}+\widetilde{\alpha}_{ij} \: ,
\label{eq:MEtensor}
\end{align}
with a pseudo-scalar $\alpha_\theta$ and a traceless tensor
$\widetilde{\alpha}_{ij}$. The latter can be further split into a traceless
symmetric part $\widetilde{\alpha}^\text{S}_{ij}$ and an antisymmetric part
$\widetilde{\alpha}^\text{A}_{ij}$. These three terms are associated with the
ME monopole, quadrupole and toroidal moments,
respectively~\cite{Spaldin2008,Thoele2018}.

Of particular interest is the pseudo-scalar part,
\begin{align}
\alpha_\theta={}&\frac{\theta}{2\pi}\frac{e^2}{h} \: ,
\end{align}
with $e>0$ denoting the elementary charge, $h$ the Planck constant, and the
dimensionless parameter $\theta$ that can be related to the axion field from
astroparticle physics~\cite{Franz2008}. It yields an isotropic coupling of
electric and magnetic fields and can thus be interpreted as a
condensed-matter realization of the axion electrodynamics~\cite{Wilczek1987,
Hehl2008,Visinelli2013}. The latter is characterized by modified Maxwell
equations that emerge from adding the term $\mathcal{L}_\text{ax}=
\alpha_\theta\;\boldsymbol{\mathcal{E}}\cdot\mathbf{B}$ to Maxwell's
Lagrangian of classical electromagnetism. The resulting modifications only
lead to physical effects if the axion field $\theta$ varies in space or time. A
spatial variation of $\theta$ arises naturally by the presence of interfaces of
ME and non-ME materials where, as a consequence, an anomalous Hall
conductivity appears~\cite{Qi2008,Essin2009}. Also, the axion field $\theta$
possesses two fundamental properties: (i) it is invariant under the shift $\theta
\rightarrow\theta+2\pi$; and (ii) due to the distinct transformation properties of
$\boldsymbol{\mathcal{E}}$ and  $\mathbf{B}$ under time reversal and spatial
inversion, the axion field has to be odd with respect to both symmetry
operations. In traditional ME media~\cite{Fiebig2005,Ramesh2007,
Spaldin2017,Fiebig2016}, the latter property is realized by the coexistence of
an intrinsic ferromagnetic and ferroelectric order. In contrast, this precondition
is circumvented in topological insulators without broken time-reversal and
inversion symmetries by the equivalence of $\theta=\pm\pi$ due to property (i).
Yet, this constrains the axion field to only appear in a quantized form, i.e.,
$\theta\in \{0,\pi\}$~\cite{Qi2011}.

It has recently been established that the ME effect is also present in bilayer
graphene and shows intriguing properties~\cite{Zuelicke2014, Winkler2015,
Zuelicke2017}. Here, the corresponding ME Lagrangian is anisotropic and
may be approximated as~\cite{Zuelicke2014}
\begin{align}
\mathcal{L}_\text{ME}=-e\; n_\mathrm{v} (\xi_\parallel\;
\boldsymbol{\mathcal{E}}_\parallel\cdot\mathbf{B}_\parallel+\xi_z\;
\boldsymbol{\mathcal{E}}_z\cdot\mathbf{B}_z)\;\delta(z) \: ,
\end{align}
where the delta distribution $\delta(z)$ locates the bilayer to be in the $x$-$y$
plane. It involves the valley-isospin density $n_\mathrm{v}$, i.e.,  the
difference of electron densities in the two ($\mathbf{K}$ and $\mathbf{K'}$)
valleys. On the one hand, a finite density imbalance is generated by the
axion-like ME coupling, as the latter induces a valley-contrasting potential
shift~\cite{Xiao2007,Nakamura2009}. At the same time, this density
dependence makes the strength of the ME response tunable and establishes
a link to valleytronics~\cite{Schaibley2016,Sui2015,Shimazaki2015}, which is
not the case in other known ME media. On the other hand, the special
transformation properties of the valleys enable the presence of the ME effect
even though time reversal and spatial inversion are symmetries of the crystal
lattice. Since this observation is general and applies to any multi-valley
system~\cite{Novoselov2016} with similar symmetries, this indicates that there
is another class of ME active materials with bilayer graphene being the first of
its kind to be discovered. Aside from this, the strength of the axion field is
determined by the material-dependent parameters, where the
two-dimensional sheet geometry suggests that there should be generally
distinct in-plane (proportional to $\xi_\parallel$) and out-of-plane (proportional
to $\xi_z$) contributions. The parameter $\xi_z$ has been evaluated in
Ref.~\onlinecite{Winkler2015}, but the value of $\xi_\parallel$ has until now
been unknown.

In this paper, we fill the knowledge gap about the ME response of a
Bernal-stacked graphene bilayer in the presence of in-plane homogeneous
electric and magnetic fields. Using a tight-binding description and applying
quasi-degenerate perturbation theory, we analytically derive an effective
low-energy Hamiltonian that comprises all relevant in-plane ME couplings and
exhibits the ME equivalence~\cite{Winkler2015}. This allows us to express the
prefactor $\xi_\parallel$ in terms of tight-binding parameters. Inserting
numerical values, $\xi_\parallel$ turns out to be approximately twice as large
as $\xi_z$ and of the same sign. Additionally, small anisotropic ME terms
induced by the electron-hole-symmetry breaking hopping $\gamma_4$ are
found. To establish a connection to experiment, we discuss the impact
of the ME couplings on features exhibited in the optical conductivity. For this
purpose, the given system configuration is particularly suitable as, in presence
of in-plane fields, the system remains metallic and the axionic response can
be more prominent than for perpendicular fields. Also, complications arising
from Landau quantization~\cite{Zhang2011} can be avoided if the magnetic field
is in-plane. We explicitly demonstrate that, due to the axionic term, the minimum
optical absorption frequencies become valley-dependent for a non-vanishing
chemical potential. Apart from this, the corrections arising from small
anisotropic ME couplings lead to a broadening of the absorption peak at zero
chemical potential. We treat the response of the graphene bilayer to the
static in-plane electric field by invoking the drift-induced Fermi-sea
displacement when calculating the optical conductivity.

This paper is structured as follows. In the following section, the general
definitions concerning the crystal lattice and tight-binding model of bilayer
graphene are briefly reviewed. In Sec.~\ref{sec:ME_Hamiltonian}, we first
derive a tight-binding Hamiltonian describing our system of interest in the
presence of an in-plane magnetic field by taking into account the arising
Peierls phases. Including in-plane electric fields, in the next step, we compute
an effective two-band Hamiltonian for the low-energy regime that contains all
relevant ME couplings. In Sec.~\ref{sec:conductivity}, we employ these results
to study the impact on the optical conductivity, which turns out to exhibit
distinctive features arising from the ME response. Electronic-structure
parameters used for numerical calculations in this paper are listed in
Table~\ref{table:parameters}.

%---------------------------------------------------------------------
\section{Basic theory for the electronic structure of bilayer graphene}

\subsection{Crystal structure}

\begin{figure}[t]%[htbp]
\includegraphics[width=.75\columnwidth]{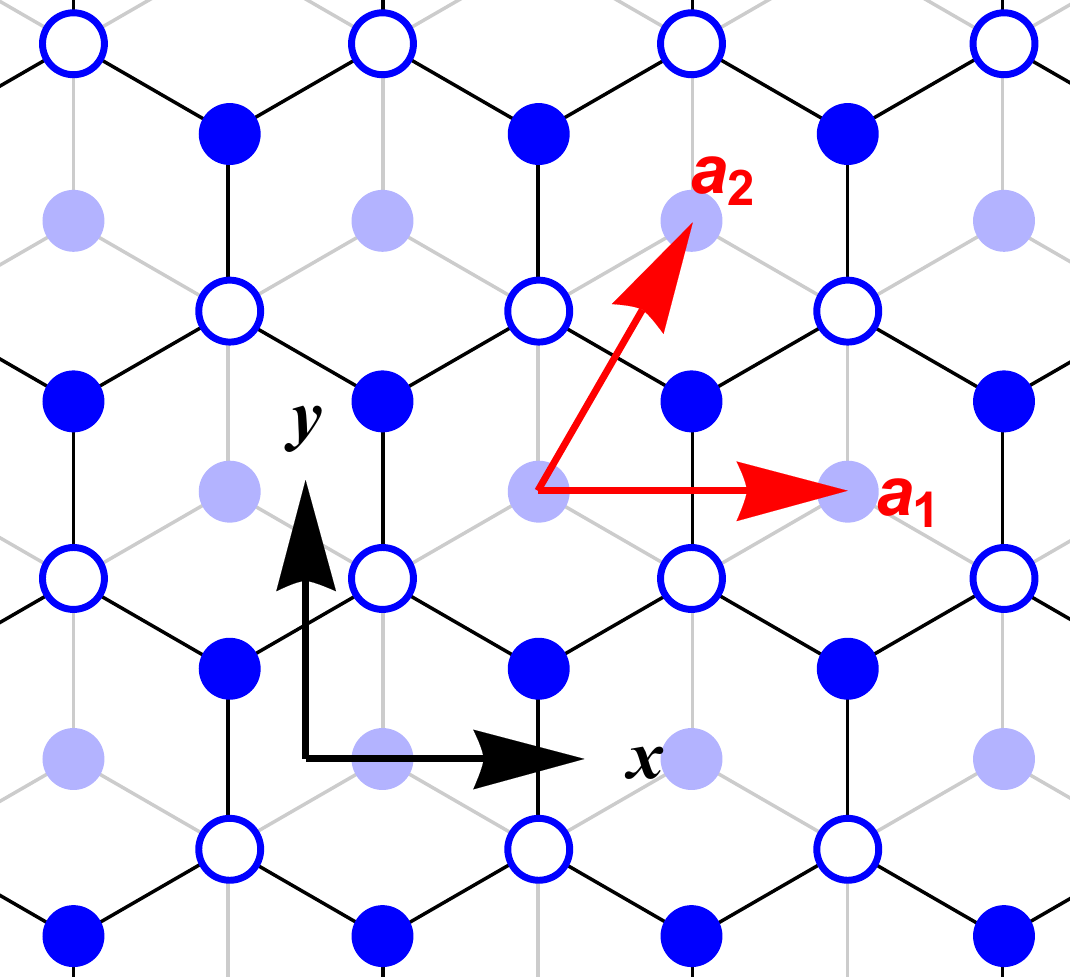}
\caption{Top view on the crystal lattice of a Bernal-stacked graphene bilayer.
Open circles indicate sites on the dimer sublattices $A$ and $A'$, while the
blue (light blue) closed circles are sites on the non-dimer sublattice $B$
($B'$).}
\label{fig:2D_lattice}
\end{figure}

The crystal lattice of a Bernal-stacked graphene bilayer is defined in
accordance with Ref.~\onlinecite{Winkler2015} as illustrated in
Figs.~\ref{fig:2D_lattice} and \ref{fig:3D_lattice}. The bilayer is composed of
two coupled graphene monolayers which are characterized in real space by
the primitive lattice vectors
\begin{align}
\mathbf{a}_1={}&a\;(1,0,0)^\top,\\
\mathbf{a}_2={}&\frac{a}{2}\;(1,\sqrt{3},0)^\top,
\end{align}
with the lattice constant $a$.
Note that the distance of two adjacent carbon atoms within each layer is $a/
\sqrt{3}$. Each of the coupled monolayers consists of two sublattices, which
we define as ($A,B$) for the top and ($A',B'$) in the bottom layer, forming a
hexagonal lattice. In the Bernal-stacked form the atoms are arranged such that
the sublattices $A$ and $A'$ lie on top of each other, i.e., they are connected
by the vector
\begin{align}
\mathbf{a}_3={}&d\;(0,0,1)^\top,
\end{align}
where $d$ is the inter-layer distance.
In contrast, the other sublattices ($B,B'$) are displaced such that the
corresponding atom is normal to the center of each hexagon of the other layer.
In other words, the top layer can be generated by shifting the bottom layer by
the vector $\mathbf{a}_3$ followed by a reflection at the $x$-$z$ plane. The
point group of the bilayer graphene is $D_{3d}$~\cite{McCann2013}. The
atomic sites ($A,A'$) are referred to as dimer and the sites ($B,B'$) as
non-dimer sites.

The intra-layer nearest-neighbor and second-nearest-neighbor vectors
$\boldsymbol{\tau}_1$ and $\boldsymbol{\tau}_2$, with respect to, e.g.,
sublattice $A$, can be written as
\begin{align}
\boldsymbol{\tau}_1^{(j)}={}&\mathcal{R}(2\pi j/3)\;\boldsymbol{\tau}_1^{(3)},
&(j\in\{1,2,3\}),\\
\boldsymbol{\tau}_2^{(j)}={}&\mathcal{R}(2\pi j/6)\;\mathbf{a}_1, &(j\in\{1,\dots,
6\}).
\end{align}
Here, $\mathbf{\tau}_1^{(3)}=a \mathbf{\hat{y}}/\sqrt{3}$ and $\mathcal{R}
(\phi)$ denotes a rotation about the $z$-axis by the angle $\phi$.

In $\mathbf{k}$-space, the lattice retains its hexagonal shape but is rotated by
$\pi/2$ about the $z$-axis. The according primitive reciprocal lattice vectors
read as
\begin{align}
\mathbf{b}_1={}&\frac{2\pi}{\sqrt{3} a}\;(\sqrt{3},-1,0)^\top,\\
\mathbf{b}_2={}&\frac{4\pi}{\sqrt{3} a}\;(0,1,0)^\top,\\
\mathbf{b}_3={}&\frac{2\pi}{d}\;(0,0,1)^\top.
\end{align}
The two sublattices give rise to two kinds of inequivalent corner points
$\mathbf{K}$ and $\mathbf{K'}\equiv - \mathbf{K}$, where
\begin{align}
\mathbf{K}={}&\frac{4\pi}{3 a}\;(1,0,0)^\top.
\end{align}
These corner points or \textit{valleys} are of fundamental interest as the
bandgap is minimal there or vanishes.

%-----------------------------------------------------------------------------
%from /Bilayer_graphene/figs/Bilayer_lattice.nb
%-----------------------------------------------------------------------------

%---------------------------------------------------------------------
\subsection{Tight-binding description}

The tight-binding model has been used to study the electronic bandstructure
for bilayer graphene in many different contexts. For a comprehensive review,
see e.g. Refs.~\onlinecite{CastroNeto2009,McCann2013,Rozhkov2016}.

Within the tight-binding approach, the eigenfunctions are linear combinations
of the Bloch functions
\begin{align}
\psi_{n,\mathbf{k}}(\mathbf{r})={}&\frac{1}{\sqrt{N}} \sum_{l=1}^{N} e^{i \mathbf{k}\cdot\mathbf{R}_l} \varphi_n(\mathbf{r}-\mathbf{R}_l),
\end{align}
where $\mathbf{k}$ denotes the (purely in-plane) wave vector of charge
carriers in bilayer graphene, $\varphi_n(\mathbf{r}-\mathbf{R}_l)$ is the $n$th
atomic orbital at the lattice site $\mathbf{R}_l$, and $N$ is the total number of
lattice sites~\cite{McCann2013}. Small corrections that arise from the
non-orthogonality of the Bloch functions shall  be neglected in the following. As
it is the anti-bonding $\pi$ bonds derived from $p_z$-orbitals that are relevant
for the electronic transport, we consider one $p_z$-orbital for each of the four
sites within the unit cell, i.e., $n \in \{A, B, A', B'\}$. With this, we represent the
tight-binding Hamiltonian $\mathcal{H}$ in the basis $\{\ket{\psi_A},
\ket{\psi_B},\ket{\psi_{A'}},\ket{\psi_{B'}}\}$, where we use this very ordering. 
Including a magnetic field $\mathbf{B}=\nabla\times \mathbf{A}(\mathbf{r})$
with the vector potential $\mathbf{A}(\mathbf{r})$, the Bloch function picks up
a Peierls phase~\cite{Kheirabadi2016,Kheirabadi2018}. A general matrix
element thus becomes
\begin{align}
\braket{\psi_m|\mathcal{H}|\psi_n}&{}\approx\notag\\
&\hspace{-.5cm}
\frac{1}{N}\sum_{l,j}\exp\left[i \mathbf{k}\cdot(\mathbf{R}_l-\mathbf{R}_j)-
\frac{i e}{\hbar}\int_{\mathbf{R}_l}^{\mathbf{R}_j}{\rm d}\mathbf{r}\cdot\mathbf{A}
(\mathbf{r})\right]\notag\\ &\hspace{-.5cm}
\times\int_{\mathbb{R}^3} {\rm d}r^3 \varphi_m^*(\mathbf{r}-\mathbf{R}_j)\,\mathcal{H}\,
\varphi_n(\mathbf{r}-\mathbf{R}_l),
\end{align}
where the vector potential yields a phase given by a line integral between the
different lattice sites.
%-----------------------------------------------------------------------------
%from /Bilayer_graphene/figs/Bilayer_lattice.nb
\begin{figure}[t]
\includegraphics[width=.75\columnwidth]{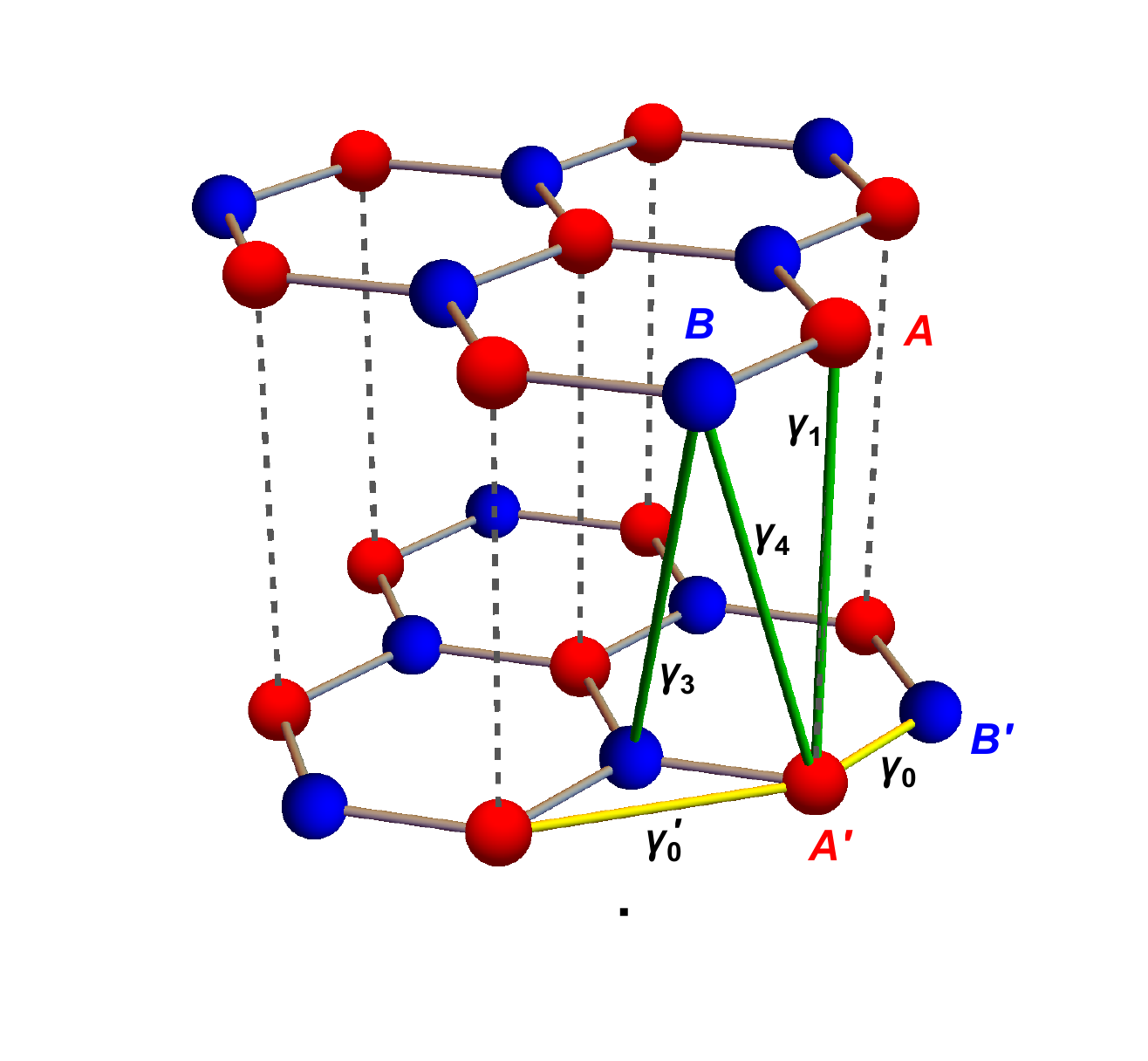}
\caption{Side view of the crystal lattice of a Bernal-stacked graphene bilayer.
The red (blue) spheres depict the atoms of the sublattices $A$ and $A'$ ($B$
and $B'$). The yellow and green connecting lines represent the intra- and
inter-layer couplings, respectively. }
\label{fig:3D_lattice}
\end{figure}
%------------------------------------------------------------------------------
The on-site energies are defined as $\epsilon_n=\braket{\psi_n|\mathcal{H}|
\psi_n}$, where we assume for simplicity $\epsilon_A=\epsilon_{A'}=
\epsilon_B=\epsilon_{B'}$, i.e., energy differences between the distinct
sublattices as well as dimer and non-dimer sites are neglected. In this paper, we
account for the couplings between nearest and second-nearest neighbors that
are characterized by the hopping integrals $\gamma_0$, $\gamma_0'$,
$\gamma_1$, $\gamma_3$, and $\gamma_4$, as schematically illustrated in
Fig.~\ref{fig:3D_lattice}. A more detailed definition is given in
Appendix~\ref{app:TBparameters}, and numerical values of all parameters are
listed in Table~\ref{table:parameters}. Typically, the  nearest-neighbor hopping
parameter $\gamma_0$ (intra-layer) and $\gamma_1$ (inter-layer) constitute
the dominant couplings. The $\gamma_3$ hopping leads to a trigonal warping
of the bandstructure, and the parameters $\gamma_4$ and $\gamma_0'$ break
the electron-hole symmetry.

For vanishing magnetic fields, the Hamiltonian takes the form~\cite{Winkler2015}
\begin{align}
\mathcal{H}&{}=
\begin{pmatrix}
\gamma_0' f_2 & -\gamma_0 f_1 & \gamma_1 & \gamma_4 f_1^* \\
-\gamma_0 f_1^*& \gamma_0' f_2 & \gamma_4 f_1^* & -\gamma_3 f_1 \\
\gamma_1 & \gamma_4 f_1 & \gamma_0' f_2 & -\gamma_0 f_1^*\\
\gamma_4 f_1 & -\gamma_3 f_1^* & -\gamma_0 f_1 & \gamma_0' f_2 
\end{pmatrix},
\end{align}
where 
\begin{align}
f_1(\mathbf{k})&{}=e^{i k_y a/\sqrt{3}}+2 e^{-i k_y a/(2\sqrt{3})}\cos(k_x a/2)
\end{align}
and $f_2(\mathbf{k})=\vert f_1(\mathbf{k})\vert^2-3$.
Expanding $f_1$ in the vicinity of the high-symmetry points $\mathbf{K}$ and
$\mathbf{K'}\equiv -\mathbf{K}$, $f_{1/2}(\mathbf{k})\mapsto f_{1/2}(\pm
\mathbf{K}+\mathbf{k})\equiv f_{1/2}^\pm(\mathbf{k})$, gives 
\begin{align}
f_1^\pm(\mathbf{k})&{}=\mp\frac{\sqrt{3}}{2}a k_\mp + \frac{a^2}{8}k_\pm^2+
\mathcal{O}(k^3),\\
f_2^\pm(\mathbf{k})&{}=-3 +\frac{3}{4}a^2 k^2+\mathcal{O}(k^3),
\end{align}
where $k_\pm=k_x\pm i k_y$ and $k=\sqrt{k_x^2+ k_y^2}$.\footnote{Notice
that, in Ref.~\onlinecite{Winkler2015}, the prefactor of $k_\pm^2$ in the
expression for $f_1^\pm(\mathbf{k})$ differs by a factor of $2$.}

%---------------------------------------------------------------------
\begin{table}[t]
\renewcommand{\arraystretch}{1.1}
\caption{Phenomenological parameters for bilayer graphene employed for numerical calculations in this paper adopted from Refs.~\onlinecite{Winkler2015,
Konschuh2012}.}
\begin{tabular*}{\columnwidth}{l @{\extracolsep{\fill}} c c r} % centered columns (4 columns)
\hline\hline 
intralayer nearest-neighbor hopping &\:$ \gamma_0$  & = & $\SI{3.0}{eV}$  \\
intralayer second-nearest-neighbor hopping &\:$\gamma_0'$  &= & $\SI{0.22}{eV}$ \\
dimer-dimer hopping &\:$\gamma_1$  &= & $\SI{0.32}{eV}$ \\
non-dimer-dimer hopping &\:$\gamma_3$  &= & $\SI{0.25}{eV}$ \\
non-dimer-non-dimer hopping &\:$\gamma_4$  &= & $\SI{0.14}{eV}$  \\
interlayer distance &\:$d$  &= & $\SI{0.335}{nm}$ \\
lattice constant &\:$a$ &= & $\SI{0.245}{nm}$  \\
\hline\hline
\end{tabular*}
\label{table:parameters} 
\end{table}
%---------------------------------------------------------------------

%-----------------------------------------------------------------
\section{Magnetoelectric coupling of in-plane electromagnetic fields}
\label{sec:ME_Hamiltonian}

Hereafter, we employ the definitions and notations of the previous section and
derive a model Hamiltonian that includes the ME couplings due to
in-plane electric and magnetic fields.

%------------------------------------------------------------------------------
\subsection{Incorporating in-plane magnetic fields into the tight-binding
Hamiltonian}

%------------------------------------------------------------------------------
%from /Bilayer_graphene/spectrum/Hamiltonian_analysis.nb
\begin{figure*}[htbp]
\includegraphics[width=.8\linewidth]{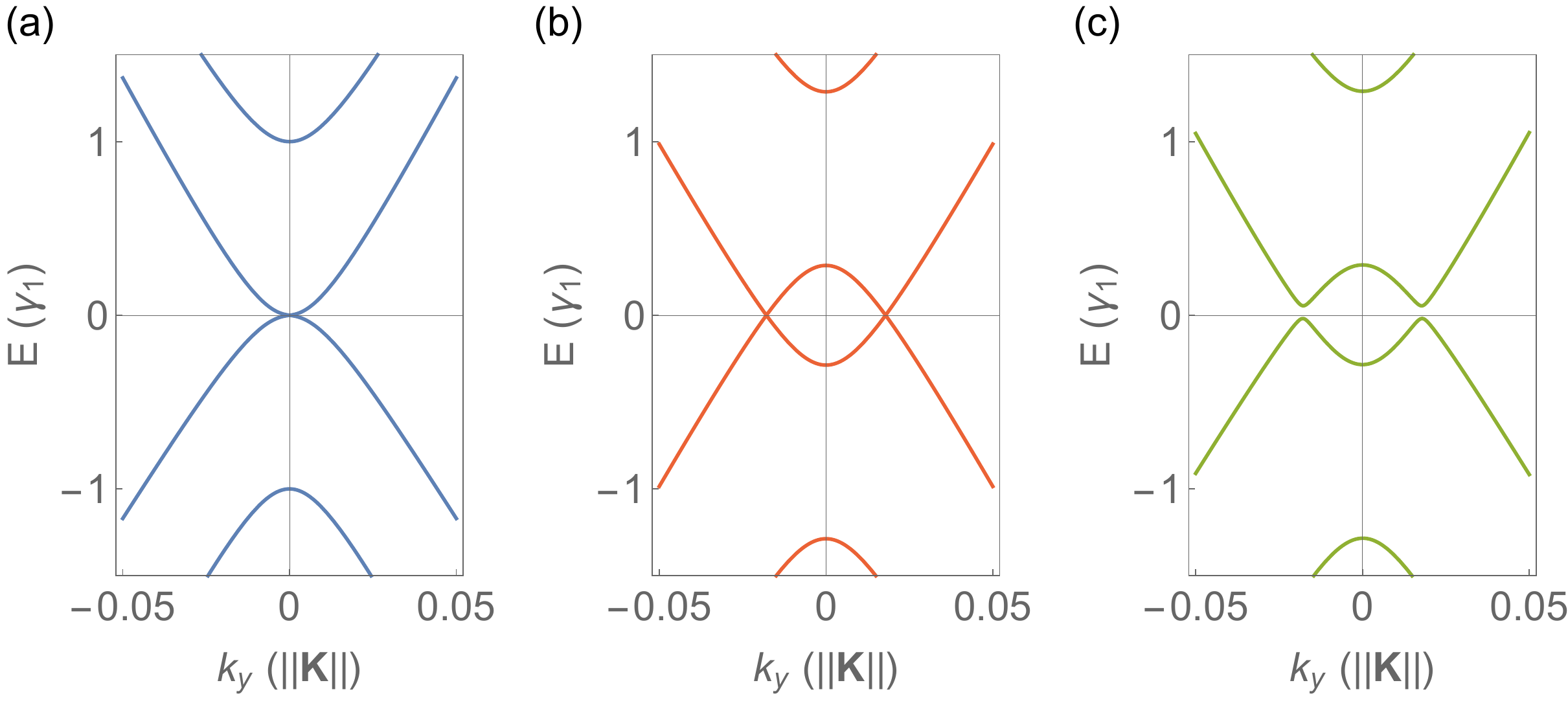}
\caption{Effect of an in-plane magnetic field $\mathbf{B} = B\,\mathbf{\hat{x}}$
on the band structure of bilayer graphene. Results shown in panel (a) [(b), (c)]
have been obtained by diagonalizing the tight-binding Hamiltonian at the
$\mathbf{K}$ valley, Eq.~(\ref{eq:Hk}), for $B=0$ [$B=\SI{1000}{T}$ and
assuming $\gamma_3=\gamma_4=\gamma_0'=0$, $B=\SI{1000}{T}$ without
any approximation]. Numerical values used for band-structure parameters are
listed in Table~\ref{table:parameters}.}
\label{fig:spectrum_B_field}
\end{figure*}
%------------------------------------------------------------------------------

The situation of an in-plane magnetic field has been addressed recently by
several authors~\cite{Pershoguba2010,Roy2013,Donck2016,Kheirabadi2016,
Kheirabadi2018}. In Ref.~\onlinecite{Pershoguba2010}, the magnetic field was
semiclassically included by adding Lorentz-force-induced momentum shifts,
and only vertical interlayer hopping $\gamma_1$ was considered. This
method was extended in Refs.~\onlinecite{Roy2013,Donck2016} by taking into
account the effect of trigonal warping. Here, we follow a more sophisticated
approach~\cite{Kheirabadi2016,Kheirabadi2018} that correctly accounts for
the arising Peierls phases in the tight-binding model.

A homogeneous in-plane magnetic field $\mathbf{B}=B_x\mathbf{\hat{x}}+B_y
\mathbf{\hat{y}}$ can be associated with a vector potential $\mathbf{A}=z(B_y
\mathbf{\hat{x}}-B_x \mathbf{\hat{y}})$. Selecting this gauge, the translation
symmetry is preserved within each layer. Further assuming a symmetric
arrangement of the top and bottom layers at $z=\pm d/2$, the Peierls phase
for the inter-layer couplings vanishes. The strong dimer-dimer coupling
$\gamma_1$ opens a gap for the $(A,A')$-like states. For low energies $\vert
E \vert<\gamma_1$, the bands of interest are described by the $(B,B')$-like
states. Hence, it is common to project on the $(B,B')$ subspace and
perturbatively include the couplings to the other bands. Following this
approach, we incorporate the effects of a small magnetic field in the
Hamiltonian; that is, we retain terms up to first (second) order in the field or the
wave vector on the off-diagonal (diagonal) part of the Hamiltonian. Since the
energy dispersion of bilayer graphene turns out to be dominated by terms that
are quadratic in the wave vector, we allow, in addition, terms quadratic in the
wave vector off-diagonal in the  $(B,B')$ sector as we project on that
subspace. Rearranging further the basis functions as $\{\frac{1}{\sqrt{2}}
(\ket{\psi_A}+\ket{\psi_{A'}}),\frac{1}{\sqrt{2}}(\ket{\psi_A}-\ket{\psi_{A'}}),
\ket{\psi_B},\ket{\psi_{B'}}\}$ and neglecting the constant energy shift due to
$f_2$, we obtain
\begin{widetext}
\begin{align}
\mathcal{H}(\pm\mathbf{K}+\mathbf{k})&{}\approx
 \renewcommand{\arraystretch}{1.5}\setlength{\arraycolsep}{9pt}
\begin{pmatrix}
%---------------------------------------------------------------------
%---------------------------------------------------------------------
\gamma_1 +\tilde{\gamma}_0' \vert \kappa_2\vert^2  &
0 &
[\tilde{\gamma}_0\kappa_2^*-\tilde{\gamma}_4 \kappa_0^*]/\sqrt{2} &
[\tilde{\gamma}_0\kappa_1-\tilde{\gamma}_4 \kappa_0 ]/\sqrt{2}   \\
%---------------------------------------------------------------------
 0 & 
 -\gamma_1 +\tilde{\gamma}_0' \vert \kappa_1\vert^2  & 
[\tilde{\gamma}_0\kappa_2^*+\tilde{\gamma}_4 \kappa_0^*]/\sqrt{2}   &
-[\tilde{\gamma}_0\kappa_1+\tilde{\gamma}_4 \kappa_0]/\sqrt{2}   \\
 %---------------------------------------------------------------------
[\tilde{\gamma}_0\kappa_2-\tilde{\gamma}_4 \kappa_0]/\sqrt{2}  &
[\tilde{\gamma}_0\kappa_2+\tilde{\gamma}_4 \kappa_0]/\sqrt{2}  &
\tilde{\gamma}_0' \vert \kappa_2\vert^2 & 
-\tilde{\gamma}_{31}  \kappa_0^*+\tilde{\gamma}_{32}\kappa_0^2 \\
 %---------------------------------------------------------------------
[\tilde{\gamma}_0\kappa_1^*-\tilde{\gamma}_4 \kappa_0^* ]/\sqrt{2}  &
-[\tilde{\gamma}_0\kappa_1^*+\tilde{\gamma}_4 \kappa_0^*]/\sqrt{2} &
-\tilde{\gamma}_{31} \kappa_0+\tilde{\gamma}_{32} (\kappa_0^*)^2 & 
\tilde{\gamma}_0' \vert \kappa_1\vert^2
%---------------------------------------------------------------------
%---------------------------------------------------------------------
\end{pmatrix},
\label{eq:Hk}
\end{align}
with
\begin{align}
\kappa_0={}\pm k_\pm ,\quad
\kappa_1={}\pm(k_\pm \pm i b_\pm),\quad
\kappa_2={}\pm(k_\pm \mp i b_\pm),\quad
b_\pm={}\frac{e d}{2\hbar}(B_x\pm i B_y),
\end{align}
and 
\begin{align}
\tilde{\gamma}_0={}\frac{\sqrt{3}a}{2}\gamma_0,\quad
\tilde{\gamma}_0'={}\frac{3a^2}{4}\gamma_0',\quad
\tilde{\gamma}_{31}={}\frac{\sqrt{3}a}{2}\gamma_3,\quad
\tilde{\gamma}_{32}={}\frac{a^2}{8}\gamma_3,\quad
\tilde{\gamma}_4={}\frac{\sqrt{3}a}{2}\gamma_4.
\end{align}
\end{widetext}
Disregarding the parabolic terms $\propto\tilde{\gamma}_{0}',
\tilde{\gamma}_{32}$, this result coincides with the Hamiltonian given in
Ref.~\onlinecite{Kheirabadi2016} apart from a unitary transformation  and  the
sign of the $\gamma_3$-terms. Without magnetic field, this expression
corresponds to the Slonczewski-Weiss-McClure
Hamiltonian~\cite{McClure1957,Slonczewski1958}.

The energy dispersion obtained for a finite in-plane magnetic field is displayed
in Fig.~\ref{fig:spectrum_B_field}. For better visualization, we used an
extraordinary high magnetic field of $\SI{1000}{T}$. In
Refs.~\onlinecite{Pershoguba2010,Roy2013,Donck2016}, it was found that a
large in-plane magnetic field produces a change in topology of the band
structure similar to the one appearing due to lateral
strain~\cite{Kruczynski2011,Son2011,Mariani2012,He2014,Daboussi2014}.
More precisely, the parabolic low-energy dispersion splits into two Dirac
cones, cf.\ Figs.~\ref{fig:spectrum_B_field}(a) and (b), where the new Dirac
points appear at wave vectors $\mathbf{k}=\pm(\boldsymbol{\hat{z}}\times
\mathbf{b})$. However, our more detailed model includes additional hopping
terms that further reduce the symmetry and result in a gapped spectrum [cf.\
Fig.~\ref{fig:spectrum_B_field}(c)]. Due to the presence of trigonal warping, the
precise dispersion near the charge-neutrality point is quite complex and
depends sensitively on the system configuration. 

%------------------------------------------------------------------------------
\subsection{Effective low-energy Hamiltonian describing in-plane
magnetoelectric couplings}\label{sec:low_energy_Ham}

An effective Hamiltonian that describes the low-energy excitations of bilayer
graphene in absence of fields was first derived in
Ref.~\onlinecite{McCann2006b}. The result was extended in
Ref.~\onlinecite{Winkler2015} to the situation where perpendicular, static, and
homogenous electric and magnetic fields are present.

In this paper, we consider both static homogeneous electric and magnetic fields
applied within the plane of the bilayer. The magnetic field is already included in
our tight-binding model. To also account for an electric field
$\boldsymbol{\mathcal{E}}=\mathcal{E}_x \mathbf{\hat{x}}+\mathcal{E}_y
\mathbf{\hat{y}}$, we add the scalar potential ${V_ \mathcal{E}(\mathbf{r})=e\,
\boldsymbol{\mathcal{E}}\cdot\mathbf{r}}$ to the Hamiltonian in
Eq.~(\ref{eq:Hk}). Using quasi-degenerate perturbation theory
(QPT)~\cite{BirPikusBook,winklerbook}, we project on the $(B,B')$ subspace.
For the partitioning of the Hamiltonian, we select the diagonal terms proportional
to the magnetic field and all off-diagonal terms as a perturbation. To third order
in perturbation theory, this procedure yields the effective two-band Hamiltonian
$\mathcal{H}_\text{eff} +V_ \mathcal{E}(\mathbf{r})$, where
\begin{widetext}
\begin{align}
\mathcal{H}_\text{eff}={}&
%------------------------------------------------------------------------------
\left\{ 
\left( \frac{2\tilde{\gamma}_0\tilde{\gamma}_4}{\gamma_1}+\tilde{\gamma}_0' \right)k^2 \tau_0
+\frac{e\tilde{\gamma}_0^2}{\gamma_1^2}\boldsymbol{\mathcal{E}}\cdot \mathbf{b} \;\tau_z
\right\}\otimes\sigma_0\notag\\
%------------------------------------------------------------------------------
&+\left\{ 
\frac{\tilde{\gamma}_0^2-\gamma_1\tilde{\gamma}_{32}+\tilde{\gamma}_4^2}{\gamma_1}(k_y^2-k_x^2)\;\tau_0
-\left[
\tilde{\gamma}_{31} k_x+e\frac{\tilde{\gamma}_0\tilde{\gamma}_4}{\gamma_1^2}(\mathcal{E}_x b_x- \mathcal{E}_y b_y)
\right]
\;\tau_z
\right\}\otimes\sigma_x\notag\\
%------------------------------------------------------------------------------
&+\left\{ 
\left[
-\tilde{\gamma}_{31}  k_y+ e\frac{\tilde{\gamma}_0\tilde{\gamma}_4}{\gamma_1^2}(\mathcal{E}_x b_y+ \mathcal{E}_y b_x)
\right]
\;\tau_0
+2\frac{\tilde{\gamma}_0^2-\gamma_1\tilde{\gamma}_{32}+\tilde{\gamma}_4^2}{\gamma_1}k_x k_y\;\tau_z
\right\}\otimes\sigma_y\notag\\
%------------------------------------------------------------------------------
&+\left\{ 2
\frac{\tilde{\gamma}_0\tilde{\gamma}_4+\tilde{\gamma}_0'\gamma_1}{\gamma_1}( b_y k_x-b_x k_y)\;\tau_0
+e\frac{\tilde{\gamma}_0^2+\tilde{\gamma}_4^2}{\gamma_1^2}(\mathcal{E}_y k_x-\mathcal{E}_x k_y)\;\tau_z
\right\}\otimes\sigma_z.
%------------------------------------------------------------------------------
\label{eq:Heff}
\end{align}
\end{widetext}
In this notation, the Pauli matrices $\sigma_{0,x,y,z}$ are associated with the
sublattice-related pseudospin degree of freedom. On the other hand, $\tau_0$
and $\tau_z$ are Pauli matrices whose basis states represent the different
valleys in the order $(\mathbf{K},\mathbf{K'})$. Notably, the effective
Hamiltonian at the  $\mathbf{K'}$ valley can be obtained from the Hamiltonian
at the $\mathbf{K}$ valley (and \textit{vice versa\/}) by a mirror reflection at the
$yz$-plane, i.e., the polar vectors map as $k_x\rightarrow -k_x$ and
$\mathcal{E}_x\rightarrow -\mathcal{E}_x$ and the axial (pseudo-)vectors as
$B_y\rightarrow -B_y$. In above expression for $\mathcal{H}_\text{eff}$, we
excluded terms that are of third order in the wave vector or of second order in
the magnetic field. (For completeness, terms quadratic in the magnetic field
are listed in Appendix~\ref{app:parabolic}. Additional ME couplings that appear
in fourth order QPT are given in Appendix~\ref{app:higherorder}. To make the
connection with results from Ref.~\onlinecite{Winkler2015}, we discuss in
Appendix~\ref{app:perp_el_field} possible couplings between in-plane magnetic
and out-of-plane electric fields.) The effective Hamiltonian Eq.~(\ref{eq:Heff})
exhibits the ME equivalence; i.e., it is form-invariant with respect to
interchanging corresponding Cartesian components of the electric and magnetic
fields~\cite{Winkler2015}.

\begin{figure}
\includegraphics[width=.8\columnwidth]{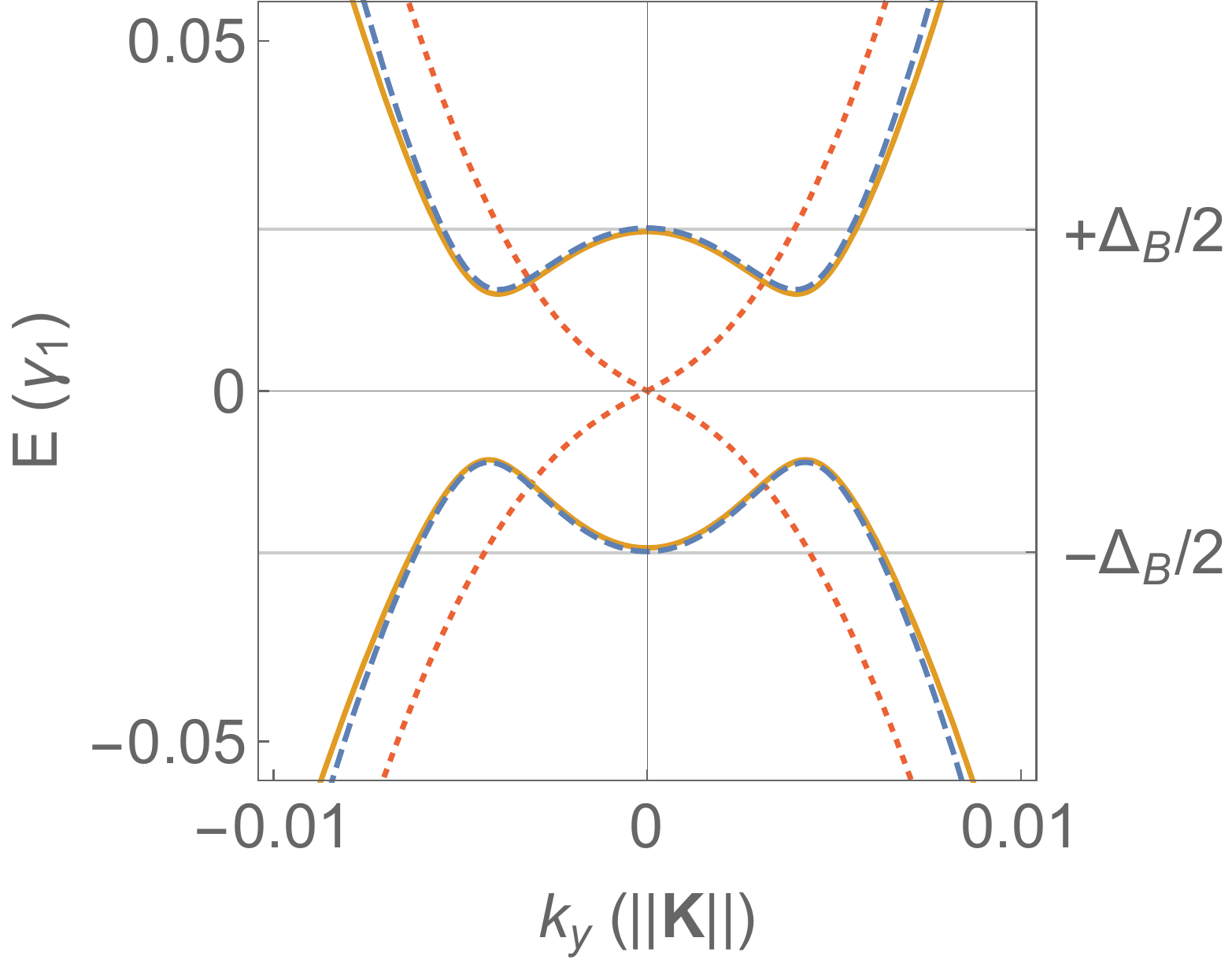}
\caption{Low-energy dispersion at the $\mathbf{K}$ valley computed for
$\boldsymbol{\mathcal{E}}=\mathbf{0}$ and $\mathbf{B}=B\, \mathbf{\hat{x}}$ with $B=
\SI{300}{T}$ using Eq.~(\ref{eq:Hk}) (yellow), Eq.~(\ref{eq:Heff}) (red-dotted),
and Eq.~(\ref{eq:Heff}) including the parabolic terms from
Eq.~(\ref{eq:B2terms}) (blue-dashed). Numerical values used for band-structure
parameters are listed in Table~\ref{table:parameters}.}
\label{fig:spectrum_model_comparison}
\end{figure}

In absence of electric fields, the magnetic-field-dependent terms in
Eq.~(\ref{eq:Heff}) appear only due to the small tunneling amplitudes
$\gamma_0'$ and $\gamma_4$ that break the electron-hole symmetry.
Hence, to observe the change of the band structure topology
\cite{Pershoguba2010,Roy2013,Donck2016} as shown in
Fig.~\ref{fig:spectrum_B_field}, we need to include corrections that are
quadratic in the magnetic field, Eq.~(\ref{eq:B2terms}). This implies that these
terms should be taken into account for large magnetic fields. On the other
hand, realistic parameters require extraordinary large magnetic fields to
observe this effect experimentally in bilayer graphene. For instance, the
characteristic energy splitting $\Delta_\text{B}$ at the $\pm\mathbf{K}$ points
can be estimated to be $\Delta_\text{B} {}= \frac{3}{8\gamma_1}\left(a d e
\gamma_0 B/\hbar\right)^2$, which yields a splitting of $\Delta_\text{B}=(
\SI{1.64e-7}{eV/T^2})\times B^2$ for bilayer graphene. A comparison of the
low-energy dispersions for a large magnetic field obtained within different
models  is provided in Fig.~\ref{fig:spectrum_model_comparison}. Ignoring the
effect of trigonal warping, the electron and hole branches have generally two
minima with a gap of approximately $4 (\tilde{\gamma}_0
\tilde{\gamma}_4/\gamma_1+ \tilde{\gamma}_0'
)\,b^2$.\footnote{As the impact of trigonal
warping is not small near the charge-neutrality point, this value largely
underestimates the real gap.} %computed in \Bilayer_graphene\Opt_conductivity\howManyMinima_B_only.nb
If both electric and magnetic fields are present, the ME and purely magnetic
terms are competing with each other. One consequence is that, above a critical
electric field strength, the electron and hole branches can have one single
extremum each. Assuming $\boldsymbol{\mathcal{E}} \parallel\mathbf{B}$ and
setting $\gamma_3=\gamma_4=\gamma_0'=0$, we can estimate the transition
to occur at $\Vert\boldsymbol{\mathcal{E}}\Vert/\Vert\mathbf{B}\Vert\approx
\gamma_1 d / (\sqrt{2} \hbar)$. Yet, we stress that the low-energy band
structure has much richer features due to the trigonal warping and the gap is
highly anisotropic. A more detailed discussion thereof goes beyond the scope of
this work and shall be presented elsewhere.

%------------------------------------------------------------------------------
%from /Bilayer_graphene/spectrum/Hamiltonian_analysis.nb
%------------------------------------------------------------------------------

Our main interest concerns the ME terms, which couple the electric and
magnetic fields and induce a breaking of the valley
degeneracy~\cite{Xiao2007,Nakamura2009,Zuelicke2014,Winkler2015}.
As the leading contribution, we identify the  axionic term
\begin{align}
\mathcal{H}_\text{ax}={}& \Delta_\text{ax}\,\tau_z \otimes \sigma_0,
\label{eq:axion}
\end{align}
with $\Delta_\text{ax}= e\, \xi_\parallel \, \boldsymbol{\mathcal{E}}\cdot
\mathbf{B}$ involving the material parameter
\begin{align}
\xi_\parallel={}&\frac{3 e d a^2}{8\hbar} \frac{\gamma_0^2}{\gamma_1^2}.
\end{align}
Using the values for tight-binding parameters given in
Table~\ref{table:parameters}, we estimate
\begin{align}
\xi_\parallel\approx{}&\SI{1.0e-3}{nm/T},
\label{eq:axion_parameter}
\end{align}
which has the same sign and is about twice as large as the prefactor for the
axionic term involving perpendicular fields; $\xi_z\approx \SI{6.0e-4}{nm/T}
$~\cite{Zuelicke2014}. Axionic terms are of particular interest as they
constitute a condensed-matter realization of axion electrodynamics. With the
values for $\xi_\parallel$ and $\xi_z$ given above, in conjunction with
expressions from Eqs.~(3a) and (8) of Ref.~\onlinecite{Zuelicke2014}, we find
the magnitude of the axionic pseudoscalar in bilayer graphene as
\begin{equation}
\theta_\mathrm{BLG} \equiv \frac{2\pi h}{e}\, \frac{2\xi_\parallel + \xi_z}{3 d}\,
n_\mathrm{v} \approx 6.7\times 10^{-3}\, n_\mathrm{v} \big[ 10^{10}
\mathrm{cm}^{-2} \big] \,\, .
\end{equation}
For comparison, the maximum value of the axionic pseudoscalar measured in
the archetypal magnetoelectric~\cite{Hehl2008} Cr$_2$O$_3$ is
$\theta_{\mathrm{Cr}_2\mathrm{O}_3}^{(\mathrm{max)}} \approx 0.13$, while in
topological insulators, a large pseudoscalar magnitude $\theta_\text{TI}=\pi$ is
fixed by time-reversal symmetry~\cite{Qi2011}. Furthermore, the axionic terms
in bilayer graphene generate an energy shift $\pm \Delta_\text{ax}$ of equal
magnitude but opposite sign for pseudospin eigenstates in the two valleys $\pm
\mathbf{K}$. This leads to a finite valley-isospin density $n_\mathrm{v}=
n_{\textbf{K}}-n_{-\textbf{K}}$, that is, the difference of charge densities $n_{\pm
\textbf{K}}$ in the distinct valleys~\cite{Xiao2007,Zuelicke2014,Winkler2015}. In
the next section, we will show how this feature manifests itself in a
valley-dependent minimal absorption frequency in the optical conductivity
spectrum when the chemical potential is not at the charge-neutrality point.

Besides the axionic contributions, Eq.~(\ref{eq:Heff}) contains other
ME-coupling terms corresponding to anisotropic contributions of the traceless
tensor $\widetilde{\alpha}_{ij}$ in Eq.~(\ref{eq:MEtensor}). They are smaller
than the uniaxial terms by a factor $\tilde{\gamma}_{4}/\tilde{\gamma}_{0}=
\gamma_{4}/\gamma_{0}\approx \SI{4.7e-2}{}$. In conjunction with the
quadratic-in-magnetic-field corrections, these terms lead to an energy gap
\begin{align}
\tilde\Delta^\pm &{}= 2 \left(\frac{\tilde{\gamma}_0}{\gamma_1}\right)^2
b \cdot\left\Vert \gamma_1 \mathbf{b}\pm
\frac{\gamma_4}{\gamma_0}e\boldsymbol{\mathcal{E}} \right\Vert \label{eq:gap}
\end{align}
between the electron and hole branches in the two valleys $\pm\mathbf{K}$.
The opposite sign of the purely ME contribution leads to a gap
difference between the valleys, i.e., $\vert\tilde\Delta^+ - \tilde\Delta^-\vert
\approx 4\gamma_4\vert\Delta_\text{ax}\vert/\gamma_0$. As shown in the
next section, this property turns out to generate a step-like structure in the
optical conductivity when the chemical potential is at the charge-neutrality point.

%---------------------------------------------------------------------
\section{Visibility of magnetoelectric coupling in the optical conductivity}
\label{sec:conductivity}

In the remainder of this paper, we explore the possibility to detect the
above-discussed ME couplings in bilayer graphene through a
transport measurement. Concomitantly with causing the ME
effects that are our primary interest, the presence of the static uniform in-plane
electric field $\boldsymbol{\mathcal{E}}$ will also generate a stationary DC
current that is associated with a shifted Fermi sea of charge carriers in bilayer
graphene. We envision applying an additional small AC electric field $\delta
\boldsymbol{\mathcal{E}}(t)$, which results in an AC contribution $\delta
\mathbf{j}(t)$ to the current density. The tensor $\sigma_{\mu\nu}(\omega)$ of
the frequency-dependent (optical) conductivity relates the Fourier-transformed
AC quantities [current density $\delta\mathbf{j}(\omega)$ and electric field
$\delta\boldsymbol{\mathcal{E}}(\omega)]$ via the linear relation ${\delta
j_\mu(\omega) = \sigma_{\mu\nu}(\omega)\, \delta\mathcal{E}_\nu(\omega)}$.
Fundamental properties of the electronic bandstructure can give rise to
distinctive features in the optical conductivity, making the latter a prime tool for
the study of unconventional materials including graphene. It has been used to
study various kinds of systems such as single or multilayer
graphene,\cite{sta08,mak08,Koshino2008,Koshino2009} bilayer graphene with
and without asymmetry gap~\cite{Nicol2008}, and  recently even taking into
account a finite twist angle between the two layers.~\cite{Stauber2018a,
Stauber2018b} In our present case, the frequency dependence of
$\sigma_{\mu\nu}(\omega)$ will not only be affected by the ME-effect-related
manipulation of the bandstructure, but also by the non-equilibrium distribution
of charge carriers within this bandstructure. In the following, we elucidate both
effects in turn. 

\subsection{Kubo formalism to calculate the frequency-dependent electric
conductivity}

We employ the Kubo formula to calculate the conductivity tensor
$ \sigma_{\mu\nu}(\omega)$. The electric-field perturbation is considered to be
spatially homogeneous, time-dependent, and parallel to the bilayer plane, i.e.,
$\delta\boldsymbol{\mathcal{E}}(t)=\delta\boldsymbol{\mathcal{E}}(\omega)
\exp [-i(\omega+i\eta)t]$. Here we have included an infinitesimally small
quantity $\eta \in \mathbbm{R}^+$. To avoid generating time-dependent
contributions to $\Delta_\mathrm{ax}$, the oscillating electric field $\delta
\boldsymbol{\mathcal{E}}(t)$ should be applied perpendicular to the static
in-plane magnetic field $\mathbf{B}$. The conductivity tensor can be
expressed in terms of the set of single-particle eigenstates $\{\ket{n}\}$ in the frequency
domain as~\cite{KammermeierPHD,wenkdiss} 
\begin{align}
\sigma_{\mu\nu}(\omega)={}&\frac{i\hbar}{\mathcal{V}}\sum_{n,l}
\frac{\braket{n|J_\mu|l}\braket{l|J_\nu|n}}{\hbar \omega+\epsilon_n-\epsilon_l
+i\eta}\frac{f(\epsilon_n)-f(\epsilon_l)}{\epsilon_l-\epsilon_n},
\label{eq:KuboConductivity1}
\end{align}
with the volume $\mathcal{V}$, the current operator $\textbf{J}=-e(\nabla_\mathbf{k}\mathcal{H})/\hbar$, and
the single-particle eigenenergies $\epsilon_{i}$. Moreover, the function
$f(\epsilon_{i})=\{1+\exp[\beta(\epsilon_i-\tilde{\mu})]\}^{-1}$, where $\beta=1/
(k_B T)$, represents the Fermi-Dirac distribution with the Boltzmann constant
$k_B$, the temperature $T$, and the chemical potential $\tilde{\mu}$. We are
interested in the dissipative part which is given by the real part of the
conductivity tensor $\text{Re}\left[ \sigma_{\mu\nu}(\omega)\right]$. Here, we
can further distinguish two terms: (i) The intra-band ($n=l$) contribution, which
determines the DC Drude conductivity, and (ii) the inter-band ($n\neq l$)
contribution, which determines the optical absorption at finite frequencies.

To illustrate the emergence of features in the optical conductivity associated
with magnetoelectricity, we now focus on the inter-band contribution to the
optical conductivity for bilayer graphene in the presence of in-plane
ME couplings in the clean limit. The low-energy bandstructure of
this system is determined by the single-particle Hamiltonian
$\mathcal{H}_\mathrm{eff}$ displayed in Eq.~(\ref{eq:Heff}). Its eigenstates
are of the form $\ket{n}=\ket{\mathbf{k}}\otimes\ket{\pm}\otimes\ket{\lambda}_\mathbf{k}$ where
$\mathbf{k}$ denotes the wave vector, $\pm$ the valley-index, and $\lambda$
distinguishes the different electron and hole branches. Since the Hamiltonian
$\mathcal{H}_\mathrm{eff}$ is diagonal with respect to $\ket{\mathbf{k}}$ and
$\ket{\pm}$, the inter-band optical conductivity of charge carriers from the
individual valleys $\pm\mathbf{K}$ becomes~\cite{Bernad2010}
\begin{widetext}
\begin{align}
\text{Re}\left[ \sigma_{\mu\nu}^\pm(\omega) \right]={}&\sigma_0\frac{\sinh
(\beta\hbar\omega/2)}{2\hbar \omega}\sum_{\lambda\neq \lambda'}\int {\rm d}
k^2 \frac{\braket{\lambda|[\nabla_\mathbf{k}\mathcal{H}_\mathrm{eff}^\pm(\mathbf{k})]_\mu|
\lambda'}\braket{\lambda'|[\nabla_\mathbf{k}\mathcal{H}_\mathrm{eff}^\pm(\mathbf{k})]_\nu|
\lambda}}{\cosh\{\beta[\epsilon_{\lambda}^\pm(\mathbf{k})+\epsilon_{
\lambda'}^\pm(\mathbf{k})-2\tilde{\mu}]/2\}+\cosh(\beta\hbar\omega/2)}\delta[\hbar \omega
-\epsilon_{\lambda'}^\pm(\mathbf{k})+\epsilon_{\lambda}^\pm(\mathbf{k})],
\label{eq:OpticalAbsorption1}
\end{align}
\end{widetext}
where $\mathcal{H}_\mathrm{eff}^\pm=\braket{\pm|\mathcal{H}_\mathrm{eff}|\pm}$,
$\sigma_0=2 e^2/h$, and we introduced an additional factor $2$ to account for
spin degeneracy. 
Here and in the following, the $\mathbf{k}$-integration is restricted to the domain around the valleys as otherwise spurious solutions may occur due to the small wave vector expansion of the Hamiltonian $\mathcal{H}_\text{eff}$. 
The total optical conductivity of the system is the sum of
contributions from the individual valleys, i.e., $\text{Re}\left[\sigma_{\mu\nu}
(\omega) \right]=\sum_\pm \text{Re}\left[\sigma_{\mu\nu}^\pm(\omega) \right]$.

In contrast to the usual situation, the distribution of charge carriers in the
unperturbed state for our case of interest is a uniformly shifted Fermi sea. See
Appendix~\ref{app:validity} for a detailed discussion. Accounting for the
stationary current-carrying state generated by the static in-plane electric field
finally amounts to using a $\mathbf{k}$-dependent chemical potential [cf.\
Eq.~(\ref{eq:mu_k})]
\begin{equation}\label{eq:BoltzCorr}
\tilde\mu(\mathbf{k}) = \tilde\mu + \frac{2 e\tilde{\gamma}_0^2}{\gamma_1}
\frac{\tau_{tr}}{\hbar}\boldsymbol{\mathcal{E}}\cdot \mathbf{k}
\end{equation}
in the expression Eq.~(\ref{eq:OpticalAbsorption1}) for the optical conductivity,
where $\tau_{tr}$ is the intra-valley transport relaxation time.

To disentangle the non-equilibrium kinetic effect of the in-plane electric field
from features associated with ME couplings, we present below results
obtained for $\sigma_{\mu\nu}(\omega)$ both with and without the
$\mathbf{k}$-dependent correction to $\tilde\mu$ included in the formula Eq.~(\ref{eq:OpticalAbsorption1}). For full consistency, life-time broadening on the
scale of $\tau_{tr}$ should also be included in the calculation of
$\sigma_{\mu\nu}(\omega)$, and the latter's salient features need to be
sufficiently separated from the intra-band (broadened-Drude-peak)
contribution to the optical conductivity to enable experimental observation.

%---------------------------------------------------------------------
\subsection{Ramification of in-plane magnetoelectric couplings for the optical
conductivity: Discussion}

In Section~\ref{sec:low_energy_Ham}, we identified two major physical
implications arising from ME effects involving in-plane electric and magnetic
fields: the valley-asymmetric axionic energy shift Eq.~(\ref{eq:axion}) and the
valley-dependent field-tunable gap Eq.~(\ref{eq:gap}). We now discuss the
features in the optical conductivity associated with each of these effects.

%------------------------------------------------------------------------------
%from /Bilayer_graphene/figs/absorption.nb
\begin{figure}
\includegraphics[width=.8\columnwidth]{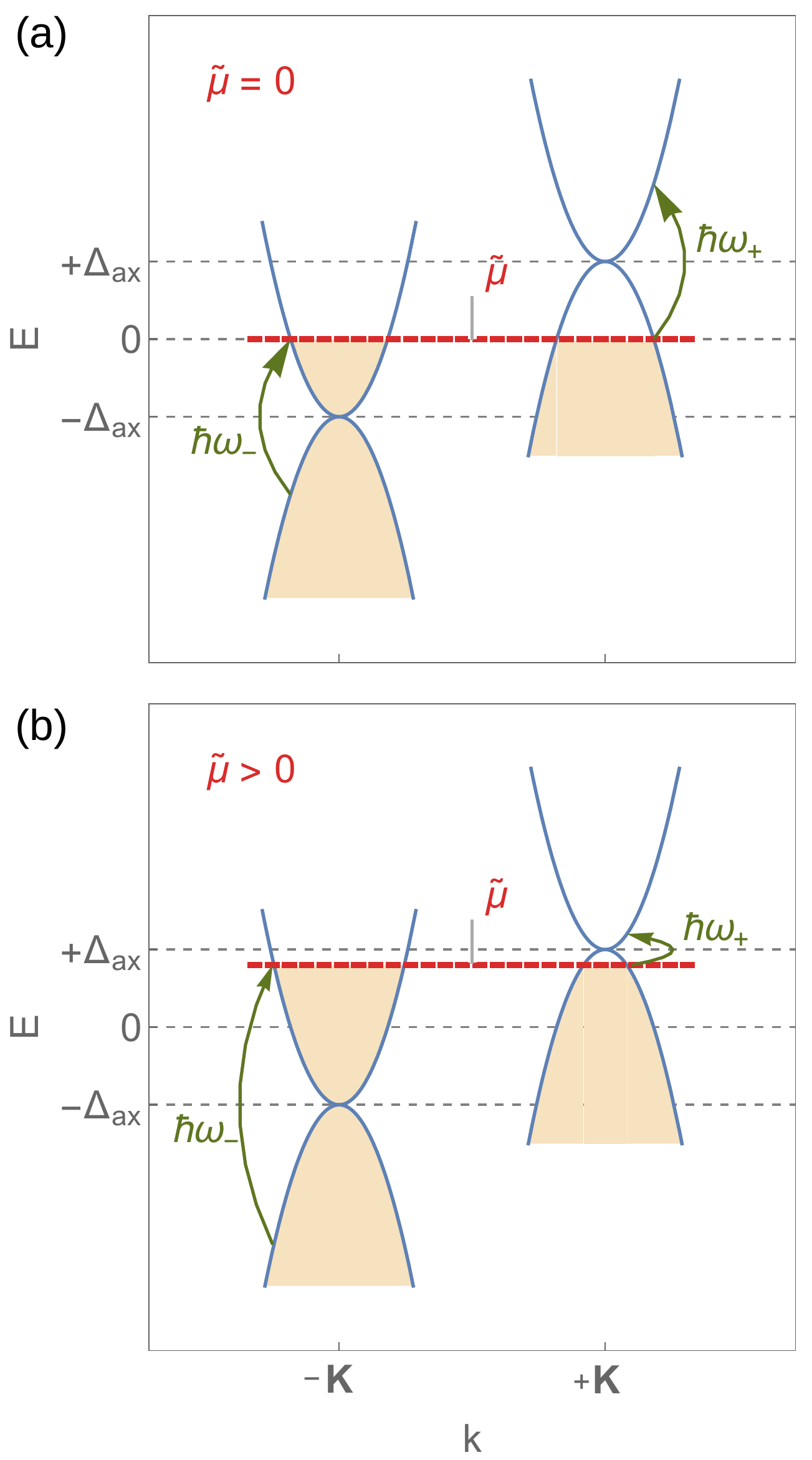}
\caption{Illustration of the valley-dependent optical absorption for different
chemical potentials $\tilde{\mu}$ due to the axionic energy shift
$\Delta_\text{ax}$. The minimum transition frequencies $\omega_\pm$ are (a)
identical for $\tilde{\mu}=0$ and (b) distinct for $\tilde{\mu}\neq 0$ (here
$\tilde{\mu}>0$).}
\label{fig:absorption}
\end{figure}
%------------------------------------------------------------------------------

%------------------------------------------------------------------------------
%from \Bilayer_graphene\Opt_conductivity\Paul\Opt_conductivity_approx_mu_shift_plot_only.nb
%and
%from \Bilayer_graphene\Opt_conductivity\Paul\Opt_conductivity_triangle_method_no_fields.nb
\begin{figure*}
\includegraphics[width=\textwidth]{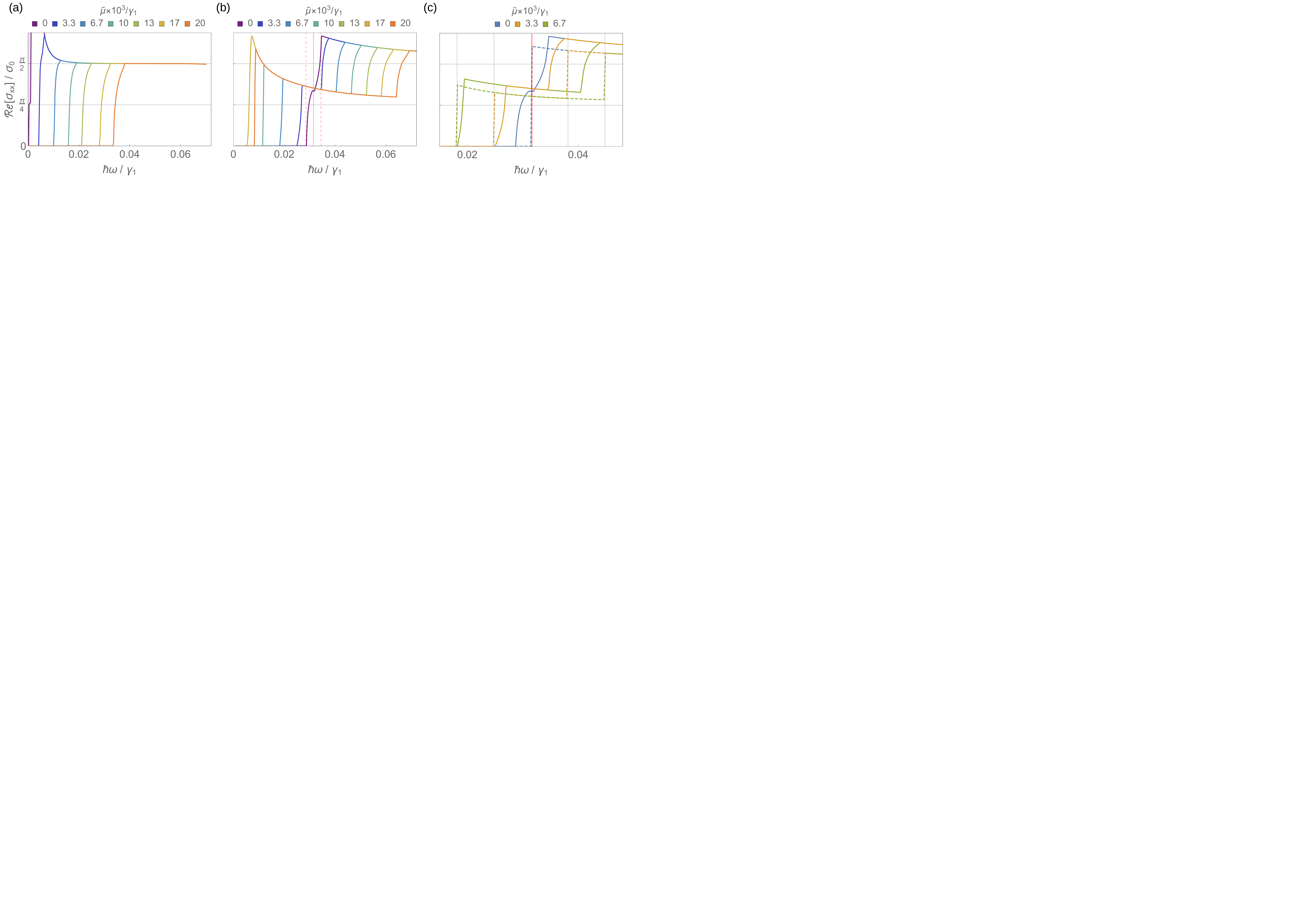}
\caption{Low-energy optical conductivity spectrum in terms of $\sigma_0=2 e^2/
h$ at zero temperature in (a) without static electric and magnetic fields and in
[(b),(c)] for  parallel fields (here, $\boldsymbol{\mathcal{E}}\parallel\mathbf{B}
\parallel\mathbf{\hat{y}}$ with $\mathcal{E}_y=\SI{0.05}{V/nm}$ and $B_y=
\SI{100}{T}$). The different colors correspond to different magnitudes of the
chemical potential $\tilde{\mu}$ in units of $10^{-3}\gamma_1$. The solid lines
correspond to the exact numerical calculation. In Fig.~(c) the plots for lowest
three chemical potentials are enlarged and compared to the simplified model
where ${\gamma_3=\gamma_4=\gamma_0'=0}$ (dashed lines). Numerical
values used for band-structure parameters are listed in
Table~\ref{table:parameters}.}
\label{fig:opt_cond}
\end{figure*}
%------------------------------------------------------------------------------

To start with, we consider the case where ${\gamma_4=\gamma_0'=0}$ and,
hence, the in-plane electric field causes only the axionic shift. [The same holds
for the in-plane magnetic field if the parabolic terms Eq.~(\ref{eq:B2terms}) are
disregarded.] Neglecting the $\mathbf{k}$-dependent correction to the chemical
potential for now, the expression Eq.~(\ref{eq:OpticalAbsorption1}) for the
conductivity tensor can be factorized into two parts,\footnote{This factorization is
generally valid~\cite{Bernad2010} for any electron-hole-symmetric
bandstructure.}
\begin{align}
\text{Re}\left[ \sigma_{\mu\nu}^\pm(\omega) \right]={}&\sigma_0\,\chi(\omega,
\beta,\tilde{\mu}_\pm)\,\Gamma_{\mu\nu}(\omega),
\label{eq:OpticalAbsorption2}
\end{align}
with valley-dependent chemical potentials
$\tilde{\mu}_\pm=\tilde{\mu} \mp \Delta_\text{ax}$ appearing only in the
function
\begin{align}
\chi(\omega,\beta,\tilde{\mu}_\pm) &{}= \frac{\sinh(\beta\hbar\omega/
2)}{\cosh(\beta\tilde{\mu}_\pm)+\cosh(\beta\hbar\omega/2)}, 
\label{eq:temp_fct} 
\end{align}
and the part
\begin{align}
\Gamma_{\mu\nu}(\omega)={}&
\sum_{\lambda\neq \lambda'}\int {\rm d}^2k\;\frac{\delta[\hbar \omega-\epsilon_{\lambda'}(\mathbf{k})+\epsilon_{\lambda}(\mathbf{k})]}{2\hbar \omega}
\notag\\&\times 
\braket{\lambda|[\nabla_\mathbf{k}\mathcal{H}_\mathrm{eff}(\mathbf{k})]_\mu|
\lambda'}\braket{\lambda'|[\nabla_\mathbf{k}\mathcal{H}_\mathrm{eff}(\mathbf{k})]_\nu|
\lambda}
\label{eq:current_correl}
\end{align}
being fully determined by the bandstructure and independent of the valley-index
which is therefore suppressed. In Fig.~\ref{fig:absorption} it is illustrated that,
depending on the value of the chemical potential $\tilde{\mu}$, the minimum
transition frequencies $\omega_\pm=2\vert\tilde{\mu}\mp\Delta_\text{ax} \vert/
\hbar$ for inter-band transitions at a fixed wave vector  are in general
valley-dependent. For $\tilde{\mu}=0$, i.e., with the chemical potential at the
charge-neutrality point, the minimal frequencies are equally large and determined by
the axionic shift. For $\tilde{\mu}\neq 0$, both peaks separate, and the
difference $\vert\omega_+-\omega_-\vert$ becomes maximal as soon as the
chemical potential exceeds the axionic energy shift, i.e., $\vert\tilde{\mu}\vert
\geq \vert\Delta_\text{ax}\vert$. As this property becomes manifest in the optical
absorption spectrum, the magnitude of the axionic term is, in principle,
accessible in experiment.

We now discuss the situation when the $\gamma_4$-terms are
included. These result in a breaking of the electron-hole symmetry and a
valley-contrasting energy gap due to additional ME couplings (cf.\
Sec.~\ref{sec:low_energy_Ham}). This leads to valley-dependent corrections
to the minimum transition frequencies, which can be approximated (by
neglecting $\gamma_3$, $\gamma_0'$, as well as quadratic terms in the
electric or magnetic field) as $\omega_\pm\approx {2(1+2\zeta\gamma_4/
\gamma_0)\vert\tilde{\mu}\mp\Delta_\text{ax} \vert/\hbar}$ where $\zeta=1$ if
$\pm\Delta_\text{ax}>\tilde{\mu}$ and $\zeta=-1$ otherwise. The effect
becomes particularly pronounced for a vanishing chemical potential where a
broadening of the minimal-frequency absorption peak of $\vert\omega_+-
\omega_-\vert=8\gamma_4\vert\Delta_\text{ax}\vert/(\gamma_0\hbar)$ occurs.

The case where we neglect electron-hole asymmetry and trigonal warping
by setting ${\gamma_3=\gamma_4=\gamma_0'=0}$ facilitates further
analytical treatment. Including the parabolic terms in the magnetic field,
Eq.~(\ref{eq:B2terms}), the energy eigenvalues read as
\begin{widetext}
\begin{align}
\epsilon_{e/h}(\mathbf{k}) {}=& \pm\frac{\tilde{\gamma}_0^2}{\gamma_1^2}
\bigg\{ \frac{e^2}{2}\mathcal{E}^2 k^2 +\gamma_1^2(b^4+k^4) +k^2\left[
2\gamma_1^2b^2\cos(2(\phi_\mathbf{b}-\phi_\mathbf{k}))-\frac{e^2}{2}
\mathcal{E}^2\cos(2(\phi_{\boldsymbol{\mathcal E}}-\phi_\mathbf{k}))\right]
\bigg\}^{1/2},
\label{eq:eval0}
\end{align}
where the upper and lower sign corresponds to the electron ($e$) and hole
($h$) branches, respectively.  Here, we introduced $\phi_\mathbf{a}$ to indicate
the polar angle that the in-plane vector $\mathbf{a}$ encloses with the $x$ axis,
and we omitted the constant axionic shift as this is absorbed into the
valley-dependent chemical potentials $\tilde\mu_\pm$. Focusing on the
longitudinal conductivity correction (selecting $\sigma_{xx}$ without loss of
generality), we can simplify Eq.~(\ref{eq:current_correl}) to
\begin{align}
\Gamma_{xx}(\omega)={}& \frac{8\tilde{\gamma}_0^8}{(\hbar\omega)^3}\int
{\rm d}^2k\;\frac{\delta[\hbar \omega-\epsilon_{e}(\mathbf{k})+\epsilon_{h}
(\mathbf{k})]}{(\hbar\omega)^2\gamma_1^4-4\tilde{\gamma}_0^4e^2
\mathcal{E}^2k^2\sin^2(\phi_{\boldsymbol{\mathcal E}}-\phi_\mathbf{k})}
\bigg\{(\hbar\omega)^2 k^2\Big[b^2\sin(2\phi_\mathbf{b}-\phi_\mathbf{k})+
k^2\sin(\phi_\mathbf{k})\Big]^2\notag\\
&+\frac{\tilde{\gamma}_0^4}{\gamma_1^4}e^2\mathcal{E}^2\Big[b^4
\sin(\phi_{\boldsymbol{\mathcal E}})-k^4\sin(\phi_{\boldsymbol{\mathcal
E}}-2\phi_\mathbf{k})+2b^2k^2\cos(2\phi_\mathbf{b}-
\phi_{\boldsymbol{\mathcal E}}-\phi_\mathbf{k})\sin(\phi_\mathbf{k})\Big]^2
\bigg\}.
\label{eq:Gamma_xx}
\end{align}
\end{widetext}

The distinctive features appearing in the optical conductivity due to the axionic
shift are illustrated in Fig.~\ref{fig:opt_cond}, both for the idealized case
${\gamma_3=\gamma_4=\gamma_0'=0}$ and for the real system at zero
temperature. For comparison, the optical conductivity spectrum without static
electric and magnetic fields is shown in Fig.~\ref{fig:opt_cond}(a). In
Figs.~\ref{fig:opt_cond}(b) and (c) we assume $\mathcal{E}_x=0$, $B_x=0$,
$\mathcal{E}_y=\SI{0.05}{V/nm}$ and $B_y=\SI{100}{T}$. For this situation, the
axion shift is $\Delta_\text{ax}=\SI{5.0}{meV}$, and the electron and hole
branches have a single extremum (cf.\ Sec.~\ref{sec:low_energy_Ham}).
Differently colored curves represent results obtained for different chemical
potentials $\tilde{\mu}$. The vertical red solid grid lines mark the minimum
transition frequency $2\Delta_\text{ax}/\hbar$ in the case of $\tilde{\mu}=0$ and
electron-hole symmetry. In Fig.~\ref{fig:opt_cond}(b) the vertical red dashed grid
lines  depict the small broadening of the minimum frequency transition peak of
$8\gamma_4\vert\Delta_\text{ax}\vert/(\gamma_0\hbar)\approx \SI{1.8}{meV}/
\hbar$ due to the electron-hole asymmetry and the ME couplings $\propto
\gamma_4$. For increasing chemical potential the minimum absorption peaks
shift to higher or lower frequencies depending on the valley index. As soon as
the chemical potential exceeds the axion shift, both peaks move simultaneously
to higher frequencies while  retaining the constant difference of approximately
$2\Delta_\text{ax}/\hbar$ (cf.\ the orange curve for $\tilde{\mu}=20\times 10^{-3}
\gamma_1$). Notice also that for both $\tilde{\mu}=13\times 10^{-3}\gamma_1$
(light green curve) and $\tilde{\mu}=17\times 10^{-3}\gamma_1$ (yellow curve)
the chemical potential lies within the gap of one valley and therefore the
respective absorption peaks are identical. 
In Fig.~\ref{fig:opt_cond}(c) the conductivity spectrum is enlarged for the
lowest three chemical potentials. Dashed curves show results obtained from
the above-discussed simplified theory where ${\gamma_3=\gamma_4=
\gamma_0'=0}$, yielding Eq.~(\ref{eq:Gamma_xx}). In this approximation, the
minimum transition peaks are sharp and occur precisely at $\omega_\pm=2
\vert\tilde{\mu}\mp \Delta_\text{ax} \vert/\hbar$ as emphasized by the vertical
gray grid lines. Lastly, we illustrate with the horizontal gray grid lines in both
plots, that asymptotically the conductivity recovers the universal value of
$\text{Re}[\sigma_{xx}]=\frac{\pi}{2} \sigma_0$ (or $\text{Re}[\sigma_{xx}^\pm]=
\frac{\pi}{4} \sigma_0$ for either valley), which is in agreement with the
established high-frequency limit without fields.\cite{Nicol2008}

%------------------------------------------------------------------------------
%from \Bilayer_graphene\Opt_conductivity\Paul\Opt_conductivity_triangle_method_drift_c.nb
\begin{figure}
\includegraphics[width=.8\columnwidth]{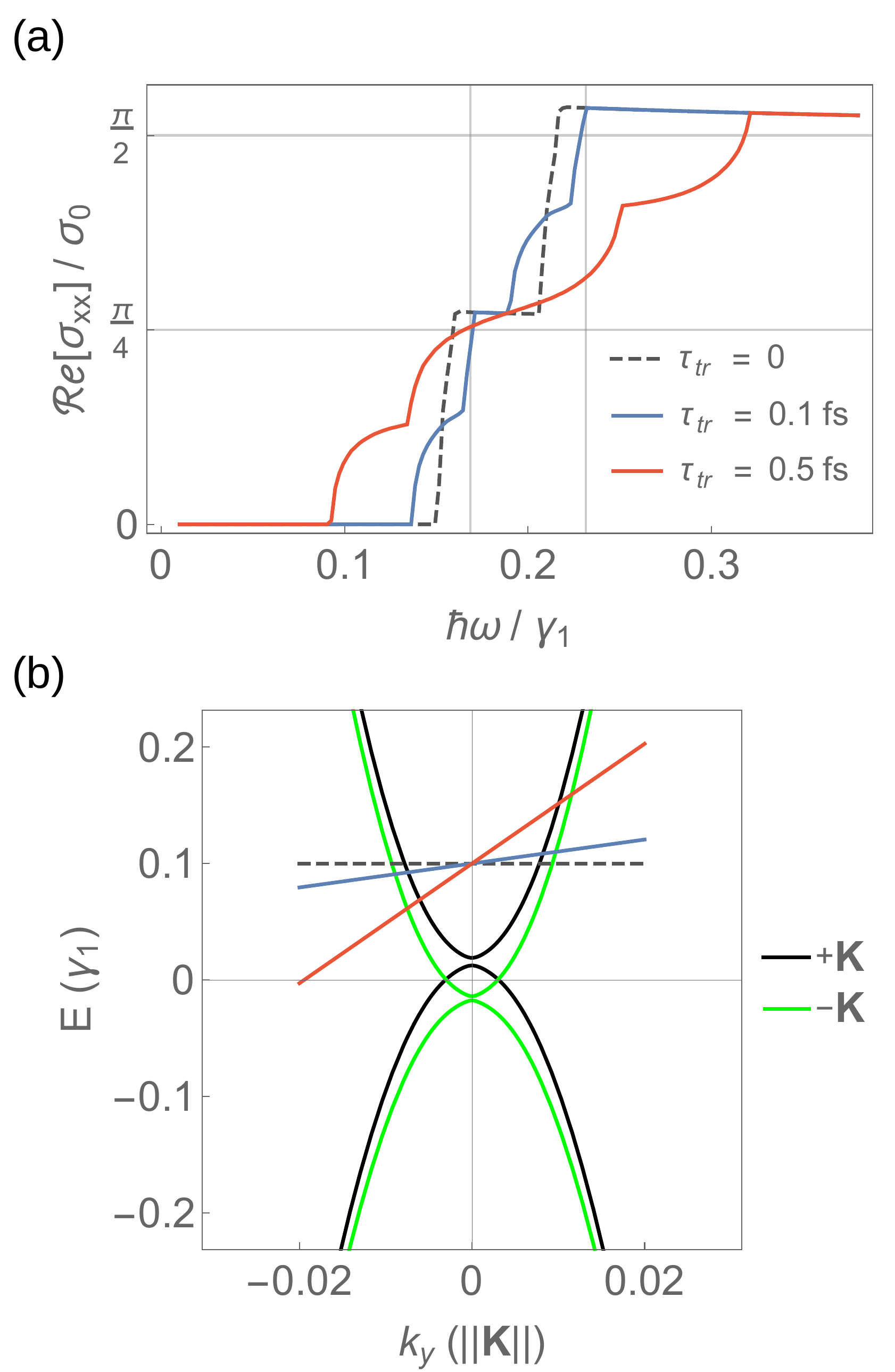}
\caption{(a) Optical conductivity in units of $\sigma_0=2 e^2/h$ obtained at zero
temperature including the corrections due to the carrier drift for different
relaxation times $\tau_{tr}$. The chemical potential at $\mathbf{k}=\mathbf{0}$ is
selected as $\tilde{\mu}(0)=0.1\gamma_1$. The fields are chosen analogously
to Fig.~\ref{fig:opt_cond}, i.e., ${\boldsymbol{\mathcal{E}}\parallel\mathbf{B}
\parallel\mathbf{\hat{y}}}$ with $\mathcal{E}_y=\SI{0.05}{V/nm}$ and $B_y=
\SI{100}{T}$. The dashed curve refers to the case without drift current. Panel (b)
shows the corresponding energy dispersion at the $\pm\mathbf{K}$ valleys for
$k_x=0$. The linear plots with different slope illustrate the
$\mathbf{k}$-dependent chemical potential $\tilde{\mu}(\mathbf{k})$,
Eq.~(\ref{eq:BoltzCorr}), for the different relaxation times. Numerical values
used for band-structure parameters are listed in Table~\ref{table:parameters}.}
\label{fig:drift_corrections}
\end{figure}
%------------------------------------------------------------------------

As discussed briefly above and in greater detail in Appendix~\ref{app:validity},
we can account for the corrections arising from the carrier drift due to the static
electric field $\boldsymbol{\mathcal{E}}$ by considering an effectively
$\mathbf{k}$-dependent chemical potential $\tilde{\mu}(\mathbf{k})$, as given in
Eq.~(\ref{eq:BoltzCorr}). The resulting modifications of the optical-conductivity
spectrum are displayed in Fig.~\ref{fig:drift_corrections}(a) for various relaxation
times and  $\tilde{\mu}(0)=0.1\gamma_1$. The vertical grid lines correspond to
the the valley-dependent minimal absorption frequencies for neglected drift
current. In Fig.\ref{fig:drift_corrections}(b) we plot the effective chemical
potential $\tilde{\mu}(\mathbf{k})$ with respect to the energy dispersion. It is
chosen large enough that the Fermi energy lies above the gap in both valleys
and the difference of the valley-dependent minimal transition frequencies is
maximal. We see that for an increasing drift, the associated energy window
obscures the ME features, albeit sharp features associated with these remain.
Further inclusion of life-time broadening into the expression
Eq.~(\ref{eq:OpticalAbsorption1}) for the optical conductivity at the assumed scale of
$\tau_{tr}$ will likely be deleterious to any such features. Yet, the issue of the
drift current can be circumvented if the magnetic field plays the dominant role in
the valley-contrasting energy shift $\Delta_\text{ax}$ (cf.\
Appendix~\ref{app:validity}).
 
While conclusive experimental detection of ME effects in pristine bilayer
graphene may be a challenge because of their smallness, the presented
illustration for their fingerprints in the optical conductivity may serve as a useful
guide to inform experimental studies focused on other two-dimensional
materials having larger ME couplings. Future work could also explore the
potential use of generally large proximity-induced exchange
fields~\cite{Wei2016} and strain-generated pseudo-magnetic
fields~\cite{Jiang2017} for boosting magnetoelectric effects in two-dimensional
materials.

%-----------------------------------------------------------------------------
\section{Conclusion and Outlook}

We have investigated the in-plane ME couplings in bilayer graphene and how
their properties give rise to distinctive features in the optical conductivity. We
developed a tight-binding description that correctly accounts for the Peierls
phases induced by a magnetic field parallel to the bilayer plane. Taking into
account second-nearest neighbor hoppings, in particular, the electron-hole
symmetry breaking contributions, it is shown that the spectrum is generally
gapped, which was not observed within the simplified models that were
employed previously.\cite{Pershoguba2010,Roy2013,Donck2016} In the next
step, we included the effect of an in-plane electric field and derived an effective
two-band Hamiltonian that comprises all relevant ME  couplings and expresses
them in terms of tight-binding parameters. We identify an axion-like
pseudoscalar contribution to the ME tensor which has the same sign and about
twice the magnitude of the previously determined out-of-plane contribution. In
addition, small corrections due to the small skew hopping amplitude
$\gamma_4$ are found, which correspond to the ME quadrupole and toroidal
moment. The Hamiltonian also exhibits the equivalence of the electric and
magnetic fields that was previously predicted by means of group-theoretical
methods.\cite{Winkler2015} A more realistic calculation should also involve
Coulomb-interaction effects. For instance, Refs.~\onlinecite{Kusminskiy2009,
Cheianov2012} showed how dielectric screening causes band-structure
renormalizations that can also be expected to influence ME couplings.

Lastly, we use the effective two-band Hamiltonian to study the impact of the ME
terms on the low-energy optical conductivity. Although in an ideal situation each
of the ME contribution yields clear features in the frequency dependence of the
conductivity, the results are obscured in bilayer graphene. This is a
consequence of the displaced Fermi contour due to a finite drift current and the
small magnitude of the axionic ME coupling in this material. Nonetheless, these
observations remain pertinent for systems with similar symmetries yet more
pronounced ME couplings or in the presence of large magnetic fields that allow
to reduce the drift-current-inducing electric fields while maintaining a significant
axionic energy shift.

%-----------------------------------------------------------------------------
\section{Acknowledgement}

This work was supported by the Marsden Fund Council from Government
funding (contract no.\ VUW1713), managed by the Royal Society Te
Ap\={a}rangi. P.W. acknowledges funding from the German Research
Foundation (DFG) via grant SFB 1277 and project 336985961.

\appendix

%---------------------------------------------------------------------
\section{Tight-binding parameters}\label{app:TBparameters}

In this paper, we take into account the (intra-layer as well as inter-layer) tunneling from each lattice site to its nearest and second-nearest neighbor as schematically depicted in Fig.~\ref{fig:3D_lattice}.
These couplings are characterized by the hopping integrals  
\begin{align}
\gamma_0={}&-\braket{\varphi_A|\mathcal{H}|\varphi_B}=-\braket{\varphi_{A'}|\mathcal{H}|\varphi_{B'}},\label{eq:hopping0}\\
\gamma_1={}&\braket{\varphi_{A'}|\mathcal{H}|\varphi_{A}},\label{eq:hopping1}\\
\gamma_3={}&\braket{\varphi_{B}|\mathcal{H}|\varphi_{B'}},\label{eq:hopping3}\\
\gamma_4={}&\braket{\varphi_{B}|\mathcal{H}|\varphi_{A'}}=\braket{\varphi_{A}|\mathcal{H}|\varphi_{B'}},\label{eq:hopping4}
\end{align}
and the second-nearest-neighbor intra-layer coupling $\gamma_0'={}\braket{\varphi_{A}|\mathcal{H}|\varphi_{A}}=\braket{\varphi_{B}|\mathcal{H}|\varphi_{B}}=\braket{\varphi_{A'}|\mathcal{H}|\varphi_{A'}}=\braket{\varphi_{B'}|\mathcal{H}|\varphi_{B'}}$.
In literature there appear various definitions for the hopping parameters with respect to their signs. 
We selected our employed definitions consistent with Ref.~\onlinecite{Winkler2015}.
Numerical values for the hopping integrals as well as other system parameters that are used in this work are listed in Table~\ref{table:parameters}.

%---------------------------------------------------------------------
\section{Parabolic correction in the magnetic field}\label{app:parabolic}

For large magnetic fields, the terms that are of second order in the magnetic field may become important as they can lead to a change of the bandstructure topology (cf. Sec.~\ref{sec:ME_Hamiltonian}).
Neglecting wave-vector dependent terms, the leading-order contribution reads as
%----------------------------------------------------------------------
\begin{align}
\mathcal{H}_\text{eff}^\mathbf{B} {}\approx &
%--------------------------------------------------------
\frac{\tilde{\gamma}_0^2}{\gamma_1}
\left[(b_y^2-b_x^2)
\tau_0 \otimes
\sigma_x +
2 b_x b_y
\tau_z \otimes
\sigma_y\right],
\label{eq:B2terms}
\end{align}
where we dropped a global shift $\propto \tau_0 \otimes \sigma_0$. Notably,
these terms are already obtained in second-order QPT.

%---------------------------------------------------------------------
\section{Higher-order magnetoelectric couplings}\label{app:higherorder}

In fourth-order QPT we obtain higher order ME couplings due to trigonal warping ($\gamma_3$) which are linear in the electric  but parabolic in the magnetic field
\begin{align}
\delta\mathcal{H}_\text{ax}^{(1)} &{}= \frac{e\tilde{\gamma}_0^2\tilde{\gamma}_{31}}{\gamma_1^3}[\mathcal{E}_y (b_x^2-b_y^2)+2\mathcal{E}_x b_x b_y] \tau_0\otimes \sigma_z,
\end{align}
or linear in both fields and the wave vector
\begin{align}
\delta\mathcal{H}_\text{ax}^{(2)} &{}= \frac{2e\tilde{\gamma}_0^2\tilde{\gamma}_{31}}{\gamma_1^3}
[
(\mathcal{E}_y b_y-\mathcal{E}_x b_x)k_x \notag\\
&\phantom{{}= \frac{2e\tilde{\gamma}_0^2\tilde{\gamma}_{31}}{\gamma_1^3}
[}
+(\mathcal{E}_x b_y+\mathcal{E}_y b_x)k_y
] 
\tau_0\otimes\sigma_0,
\end{align}
where we neglected in the latter equation subordinate terms $\propto \gamma_3\gamma_4$.

%---------------------------------------------------------------------
\section{Leading magnetoelectric couplings in presence of perpendicular electric fields}\label{app:perp_el_field}

To additionally account for perpendicular electric fields which open up a pseudospin gap,\cite{McCann2006c,Ohta2006,Zhang2009} we may include in Eq.~(\ref{eq:Hk}) the potential\cite{Winkler2015,Konschuh2012}
\begin{align}
V_\mathcal{E}^\perp &{}= \mathcal{E}_z
\begin{pmatrix}
0 &\varepsilon_{12}  &0 &0 \\
\varepsilon_{12}   &0 &0 &0 \\
0 &0 &\varepsilon_{33}  &0 \\
0 &0 &0 &-\varepsilon_{33} 
\end{pmatrix}\label{eq:perp_field},
\end{align}
where the \textit{ab initio} calculations in Ref.~\onlinecite{Konschuh2012} yield $\varepsilon_{12}=\varepsilon_{33}=0.048e\,\text{nm}$.
From symmetry considerations, this potential should give rise to ME couplings $\propto (\mathcal{E}_z B_y \sigma_x- \mathcal{E}_z B_x \sigma_y)$.\cite{Winkler2015}
Within our model, however, these terms do not appear, meaning that their prefactor vanishes exactly.
Instead, the first non-vanishing terms are obtained in third-order QPT and correspond to higher ME couplings analogously to the in-plane couplings in Appendix~\ref{app:higherorder}.
These terms read as
\begin{align}
\delta\mathcal{H}_\text{ax}^\perp {}= 
&{}\frac{2 \tilde{\gamma}_0^2( \varepsilon_{12}+ \varepsilon_{33})}{\gamma_1^2}\mathcal{E}_z(k_y b_x-k_xb_y)\tau_0\otimes\sigma_0\notag \\
&{}+\frac{2 \tilde{\gamma}_0\tilde{\gamma}_4\varepsilon_{12}}{\gamma_1^2} 
\mathcal{E}_z
\big[(k_x b_y+k_y b_x)\tau_0\otimes\sigma_x\notag\\
&{}\phantom{+\frac{2 \tilde{\gamma}_0\tilde{\gamma}_4\varepsilon_{12}}{\gamma_1^2} 
\mathcal{E}_z
\big[}
+(k_xb_x-k_yb_y) \tau_z\otimes\sigma_y
\big] \notag \\
&{}
-\frac{ \tilde{\gamma}_0^2( \varepsilon_{12}+ \varepsilon_{33})}{\gamma_1^2}\mathcal{E}_z b^2\tau_0\otimes\sigma_z.
\end{align}
Hence, the mixing  of electric and magnetic fields perpendicular to each other causes only small corrections to the axionic terms.

%---------------------------------------------------------------------
\section{Preconditions for the linear response model}\label{app:validity}

In this section, we specify the requirements that ensure the validity of the linear response model for the optical conductivity in Sec.~\ref{sec:conductivity}.

Firstly, the static electric field $\boldsymbol{\mathcal{E}}$ induces a finite parallel drift of the carriers and the system is out of equilibrium.
This yields a modification of the Fermi-Dirac distribution as\cite{RammerBook} 
\begin{align}
f[\epsilon_{e/h}(\mathbf{k}),\tilde{\mu}]\rightarrow & f[\epsilon_{e/h}(\mathbf{k}\mp e \boldsymbol{\mathcal{E}} \tau_{tr}/\hbar),\tilde{\mu}],
\end{align}
where $\tau_{tr}$ denotes the intra-valley transport relaxation time of the carriers and the upper and lower sign corresponds to the electron ($e$) and hole ($h$) branches, respectively.
Using a parabolic approximation for the spectrum $\epsilon_{e/h}(\mathbf{k})\approx\pm(\tilde{\gamma}_0^2 /\gamma_1) k^2+const.$, we find
\begin{align}
\epsilon_{e/h}(\mathbf{k}\mp e \boldsymbol{\mathcal{E}} \tau_{tr}/\hbar)\approx &{}\, \epsilon_{e/h}(\mathbf{k})-\frac{2 e\tilde{\gamma}_0^2}{\gamma_1}\frac{\tau_{tr}}{\hbar}\boldsymbol{\mathcal{E}}\cdot \mathbf{k}.
\end{align}
Hence, we may interpret this alteration as a correction to the chemical potential and replace for convenience in Sec.~\ref{sec:conductivity} $f[\epsilon_{e/h}(\mathbf{k}),\tilde{\mu}]\rightarrow  f[\epsilon_{e/h}(\mathbf{k}),\tilde{\mu}(\mathbf{k})]$ where we substituted
\begin{align}
\tilde{\mu}\rightarrow &{}\tilde{\mu}(\mathbf{k})=\tilde{\mu}+
\frac{2 e\tilde{\gamma}_0^2}{\gamma_1}\frac{\tau_{tr}}{\hbar}\boldsymbol{\mathcal{E}}\cdot \mathbf{k}.
\label{eq:mu_k}
\end{align}
Apparently, the displacement of the Fermi contours affects the sharpness of the absorption peaks.
To resolve the splitting of the valley-dependent absorption peaks, it is  required that
\begin{align}
\Delta_\text{ax}\gg
\frac{2 e\tilde{\gamma}_0^2}{\gamma_1}\frac{\tau_{tr}}{\hbar}\vert\boldsymbol{\mathcal{E}}\cdot \mathbf{k}\vert,
\end{align}
where the relevant wave vectors are to be evaluated at the Fermi energy, i.e., $\vert\boldsymbol{\mathcal{E}}\cdot \mathbf{k}\vert\leq \mathcal{E} k_F$. 
It is particularly illuminating to relate the above condition to the transport mean free path $l_{tr}=v_F \tau_{tr}$  with the Fermi velocity $v_F=2\tilde{\gamma}_0^2 k_F/(\gamma_1\hbar)$ as it simplifies to
\begin{align}
\Delta_\text{ax}\gg
e \mathcal{E} l_{tr}.
\end{align}
As at the same time $\Delta_\text{ax}$ is proportional $\mathcal{E}$, the condition is independent of the magnitude of the static electric field.
Selecting  $\mathbf{B} \parallel \boldsymbol{\mathcal{E}}$ and inserting the numerical value of $\xi_\parallel$  for bilayer graphene, Eq.~(\ref{eq:axion_parameter}), we may further estimate
\begin{align}
B \gg
l_{tr} /\xi_\parallel \approx (\SI{e3}{T}) \times\left(\frac{l_{tr}}{\text{nm}}\right).
\end{align}
Obviously, due to the smallness of $\xi_\parallel$ in bilayer graphene this condition is hardly fulfilled  for realistic values of the magnetic field and the mean free path.

Secondly, it should be noted that for low chemical potentials, the Fermi energy may lie very close to the charge-neutrality point where the role of disorder is generally more delicate.\cite{Li2016}
In the context of optical conductivity, one should account for the broadening $\sim\hbar/\tau_{tr}$ of the Drude peak at $\omega=0$.
Hence, the minimum absorption peaks should occur at $\vert\omega_\pm\vert\gg1/\tau_{tr}$ which may be controlled by choosing an appropriate chemical potential and the strength of the axion shift.

Thirdly, to apply the linear response theory, the oscillatory external electric field $\delta\boldsymbol{\mathcal{E}}(t)$ with amplitude $\delta\mathcal{E}$ should yield a small perturbation to the kinetic energy.
In other words, the changes of the Fermi wave vector $\delta k_F$ due to the minimal coupling to the electric field $\delta k_F=e \delta A/\hbar$, where $\delta\boldsymbol{\mathcal{E}}=-\partial_t(\delta\mathbf{A})$ yields $\delta A=-\delta{\mathcal{E}}/\omega$, should be small, i.e., $\delta k_F/k_F\ll 1$.
Therefore, we obtain the condition  
\begin{align}
\delta\mathcal{E}&{}\ll \frac{\hbar\omega}{e}k_F.
\label{eq:condition1a}
\end{align}
Using again a parabolic approximation and disregarding the drift corrections due to the static field $\boldsymbol{\mathcal{E}}$, the Fermi wave vector $k_F$ obeys the relation $k_F\approx \sqrt{\gamma_1\vert\tilde{\mu}\mp\Delta_\text{ax}\vert}/\tilde{\gamma}_0$ in the valley $\pm\mathbf{K}$.
Thus, the magnitude of the oscillatory external electric field $\delta\boldsymbol{\mathcal{E}}(t)$ should be chosen according to this condition.
For larger fields, higher-order terms in the Kubo formula should be taken into account.

%---------------------------------------------------------------------
%---------------------------------------------------------------------
%---------------------------------------------------------------------
\bibliographystyle{apsrev4-1}
\bibliography{MK}
\end{document}